\documentclass[acmsmall]{acmart}

\usepackage{amsmath,amsfonts}
\usepackage{algorithmic}
\usepackage{graphicx}
\usepackage{url}
\usepackage[ruled,vlined]{algorithm2e}
\usepackage{textcomp}
\usepackage{rotating}

\usepackage{xcolor}
\usepackage[normalem]{ulem}
\def\BibTeX{{\rm B\kern-.05em{\sc i\kern-.025em b}\kern-.08em
    T\kern-.1667em\lower.7ex\hbox{E}\kern-.125emX}}
\usepackage{pifont}

\newtheorem{definition}{Definition}

\newcommand{\cmark}{\ding{51}} 
\newcommand{\xmark}{\ding{55}} 
\AtBeginDocument{%
  \providecommand\BibTeX{{%
    Bib\TeX}}}

\setcopyright{cc}
\setcctype{by}
\copyrightyear{2025}

\acmJournal{TWEB}
\acmYear{2025} 
\acmVolume{1} 
\acmNumber{1} 
\acmArticle{1} 
\acmMonth{1} 
\acmPrice{}
\acmDOI{10.1145/3774629}

\begin{document}

\title{GPoS: Geospatially-aware Proof of Stake}

\author{Shashank Motepalli}
\affiliation{%
  \institution{University of Toronto}
  \city{Toronto}
  \state{Ontario}
  \country{Canada}}
\email{shashank.motepalli@mail.utoronto.ca}

\author{Naman Garg}
\affiliation{%
  \institution{IIIT-Delhi}
  \city{New Delhi}
  \country{India}}
\email{naman21171@iiitd.ac.in}

\author{Gengrui Zhang}
\affiliation{%
  \institution{Concordia University}
  \city{Montreal}
  \state{Quebec}
  \country{Canada}}
\email{gengrui.zhang@concordia.ca}

\author{Hans-Arno Jacobsen}
\affiliation{%
  \institution{University of Toronto}
  \city{Toronto}
  \state{Ontario}
  \country{Canada}}
\email{jacobsen@eecg.toronto.edu}

\renewcommand{\shortauthors}{Motepalli, Garg, Zhang, and Jacobsen}
\begin{abstract}
Geospatial decentralization is essential for blockchains, ensuring regulatory resilience, robustness, and fairness. We empirically analyze five major Proof of Stake (PoS) blockchains—Aptos, Avalanche, Ethereum, Solana, and Sui—revealing that a few geographic regions dominate consensus voting power, resulting in limited geospatial decentralization. To address this, we propose Geospatially-aware Proof of Stake (GPoS), which integrates geospatial diversity with stake-based voting power. Experimental evaluation demonstrates an average 45\% improvement in geospatial decentralization, as measured by the Gini coefficient of Eigenvector centrality, while incurring minimal performance overhead in BFT protocols, including HotStuff and CometBFT. These results demonstrate that GPoS can improve geospatial decentralization {while, in our experiments, incurring minimal overhead} to consensus performance.
\end{abstract}



\maketitle

\section{Introduction}
While decentralization is a core premise for effectively and robustly operating blockchain systems~\cite{nakamoto2008bitcoin}, one critical dimension contributing to decentralization—the \textit{geospatial decentralization}— remains overlooked. Geospatial decentralization refers to the geospatial distribution of validators participating in the blockchain consensus mechanisms~\cite{motepalli2023analyzing}. Geospatial centralization, i.e., clustering validators in certain regions, not only undermines decentralization but also increases vulnerability to localized risks, such as regulatory interventions, natural disasters, or targeted attacks. Key reasons for prioritizing geospatial decentralization include:
\begin{enumerate}
    \item Regulatory and Political Control. Geospatial centralization makes blockchains vulnerable to regulatory control, where governments or authorities in specific regions may exert control over the blockchain. For example, the U.S. SEC has asserted regulatory jurisdiction over Ethereum transactions based on validator locations~\cite{SECEthereum}. Furthermore, US Treasury sanctions on certain blockchain addresses raise concerns about censorship~\cite{ofac_sanctions_list,censorshipEthereum} and control by authoritarian regimes, threatening the neutrality and governance of a blockchain.
    \item Robustness Against Attacks and Failures: A geographically centralized blockchain network is vulnerable to region-specific failures such as natural disasters, cyberattacks, or geopolitical unrest. In 2021, an outage in an AWS data center impacted the liveness of the Solana blockchain~\cite{amazonSolana}. Additionally, centralized cloud providers can disable validators through policy changes~\cite{HetznerSolana}, potentially halting consensus and concentrating power in centralized entities.
    \item Equitable Participation and Fairness: Geospatial centralization provides latency advantages to validators in certain regions, enabling them to profit from front-running and maximal extractable value (MEV) opportunities~\cite{bahrani2024centralization,daian2020flash,heimbach2023ethereum}. This proximity-based advantage also promotes high-frequency trading and arbitrage~\cite{gupta2023centralizing,lewis2014flash}, concentrating control in certain regions, skewing incentives, and increasing disparities in access to consensus.
\end{enumerate}

 The problem this paper addresses is to drive a system towards supporting geospatial decentralization in blockchain consensus mechanisms. Widely adopted PoS (Proof of Stake) systems determine voting power in consensus based solely on staked assets, overlooking the geospatial distribution of validators. Moreover, existing decentralization metrics, such as the Nakamoto coefficient~\cite{motepalli2024does,balajidecentralization}, fail to account for this dimension, resulting in geospatial centralization. This compromises network resilience, enables MEV exploitation~\cite{daian2020flash}, and threatens both blockchain neutrality and equitable global participation. Addressing this problem is challenging due to the inherent trade-offs between improving geospatial decentralization and maintaining system performance, measured in throughput and latency.

 To address this challenge, we design geospatially-aware consensus mechanisms. We begin by defining our system model (Section~\ref{sec:system-model}). Then, we collect empirical data on validator stake and geospatial coordinates from leading blockchains, such as Aptos, Avalanche, Ethereum, Solana, and Sui (Section~\ref{sec:dataprocessing}). We conduct empirical analysis to quantify geospatial decentralization using the Gini coefficient of the eigenvector centrality measure. Our findings indicate significant geospatial centralization, underscoring the need for consensus mechanisms founded on more robust decentralization principles.

To enhance geospatial decentralization, we propose GPoS, a mechanism that incorporates both staked assets and geospatial distribution into voting power for consensus (Section~\ref{sec:gpos}). Using our collected data, we evaluate GPoS against traditional mechanisms, demonstrating average improvements of 45\% in geospatial decentralization (Section~\ref{sec:empirical}). We emulate validator distributions across various consensus mechanisms, including CometBFT (formerly known as Tendermint~\cite{buchman2016tendermint}) and HotStuff~\cite{yin2019hotstuff}, to show that GPoS incurs minimal performance overhead, measured by throughput and latency (Section~\ref{sec:experimental}). 

The contributions of this paper are four-fold: 
\begin{enumerate}
    \item We propose \textit{GPoS}, a geospatially-aware extension to stake-based voting power, enhancing decentralization in blockchain consensus.
    \item We collect and analyze validator geospatial and stake data from five major blockchains, creating a comprehensive dataset that facilitates reproducibility and advances decentralization research.\footnote{The dataset, including raw validator geolocations, stakes, and scripts for pre-processing (e.g., proximity merging and stake aggregation), is available in the repo: \url{https://github.com/GeoDecConsensus/geo-analysis.}}
    \item We introduce a new metric to quantify geospatial decentralization using real-world data, providing insights into validator concentration and distribution.
    \item We are among the first to empirically explore trade-offs between geospatial decentralization and performance, providing guidelines to optimize blockchain efficiency and robustness.
\end{enumerate}

We frame robustness via the standard correlated failure model: geographic concentration increases shared-fate risks (e.g., outages, policy actions). Dispersing voting power across regions and providers mitigates these vulnerabilities.

\section{System Model}
\label{sec:system-model}
Let \(\mathcal{V} = \{v_1, v_2, \ldots, v_n\}\) be the set of validators for blockchain \(C\) at epoch \(t\), where epoch \(t\) spans approximately one day. Let \(sk_{i}\) represent the secret key for validator \(v_i\), with its corresponding signature denoted as \(\text{sig}(sk_{i})\). We posit that cryptographic signature schemes are secure and robust, ensuring their resilience against known attack vectors~\cite{diffie2022new,rivest1978method}. The voting power of validator \(v_i\) in the blockchain is denoted by \(\rho_i\), where \(0 < \rho_i \leq 1\). The total voting power of the blockchain \(C\) is the sum of the voting powers of all validators, indicated by \(\sum_{v_i \in \mathcal{V}} \rho_i = 1\).

\subsection{Proof of Stake (PoS)} Blockchains are susceptible to Sybil attacks~\cite{douceur2002sybil}, where a malicious actor undermines the system by masquerading as multiple validators to gain disproportionate voting power. To mitigate this risk, many blockchains use a Proof of Stake (PoS) mechanism. In PoS, voting power is proportional to the number of (native) tokens staked, referred to as \textit{stake}. This model is secure due to the finite token supply, indicating a vested interest in blockchain security. {Let \(S_i > 0\) represent the stake of validator \(v_i\) in the blockchain \(C\), then normalized stake is represented by \(s_i > 0\), expressed as:}
\begin{equation}
\label{eq:pos1}
   s_i = \frac{S_i}{\sum_{\forall v_k \in \mathcal{V}} S_k}
\end{equation}

{The voting power of validator $v_i$, in PoS, is its normalized stake, $\rho_i = s_i$.} While the rest of the paper discusses PoS, the concepts are applicable to Delegated Proof of Stake (DPoS) systems, where stakeholders can delegate their stake to other validators.

\subsection{Weighted Consensus} The effectiveness of PoS blockchains relies on a robust consensus mechanism, which is essential for validators to agree on the blockchain's state. Two key properties define consensus: \textit{liveness}, which ensures progress by updating state, and \textit{safety}, which guarantees that all correct validators see the same state~\cite{castro1999practical}. \textit{Finality} is achieved when updated state cannot be tampered with and is irreversible.

While some blockchains, such as Bitcoin~\cite{nakamoto2008bitcoin}, prioritize liveness with eventual probabilistic finality, our study focuses on systems emphasizing instant absolute finality, prioritizing safety over liveness~\cite{lewis2023byzantine,motepalli2024does}. This approach aligns with classical Byzantine Fault Tolerance (BFT) literature~\cite{zhang2024reaching} and is exemplified by blockchains like Cosmos, where consensus finality is achieved when a \textit{quorum} of validators agrees on the transaction order and content~\cite{castro1999practical}.

A quorum, denoted as \(\mathbb{Q}\), is defined in PoS blockchains as at least two-thirds of the total voting power~\cite{motepalli2024does}, expressed as:
\begin{equation}
    \mathbb{Q} = \{ \text{sig}(s_{ki}) \, | \, V_\mathbb{Q} \subseteq \mathcal{V} \text{ and } \sum_{v_i \in V_\mathbb{Q}} \rho_i \geq \frac{2}{3} \}
\end{equation}

Here, \(\mathbb{Q}\) represents the set of signatures from validators \(V_\mathbb{Q}\) such that the total voting power of this subset meets or exceeds two-thirds of the total voting power. We assume that no more than one-third of the total voting power can be malicious, consistent with BFT assumptions~\cite{castro1999practical}. 

\subsection{Geospatial Distance} Our research examines the impact of the geospatial distribution of validators on consensus. Each validator \(v_i\) is located at coordinates \(c_i(x_i, y_i)\), where \(x_i\) represents latitude and \(y_i\) represents longitude. The distance between two validators \(v_i\) and \(v_j\) is denoted by \(\Delta_{i,j} = \text{haversine}(c_i, c_j)\), expressed in kilometers. The haversine distance is employed to accurately measure distances over the Earth's spherical geometry~\cite{sinnott1984virtues}. It is assumed that reliable network communication exists among all non-malicious validators within the system. 

\subsection{{Location Attestation and Trust Model}} In this study, we examine global geospatial trends rather than local variations, such as clustering of validators in data centers within metropolitan areas. We rely on the accuracy of location data sourced from our datasets and IP-geolocation services. 

{We acknowledge that GPoS creates economic incentives for validators to misrepresent their locations to gain greater voting power. Therefore, the security of our GPoS relies on a reasonably accurate external location attestation mechanism. Our work is not a new Proof-of-Location (PoL) protocol; rather, GPoS is a consensus-layer mechanism designed to be composable with existing or future PoL systems~\cite{sheng2024bft} that can provide verified coordinates.} Existing literature provides techniques for determining geolocation, such as topology-based methods that leverage ping latencies and distance computations to ascertain validator locations more accurately~\cite{gueye2004constraint,katz2006towards,padmanabhan2001investigation}. Although these methodologies are complementary to our solution, their specifics fall outside the current scope of this work. This study relies on the assumption of location accuracy to enhance geospatial decentralization in blockchains.

\section{Validator Data Collection and Pre-Processing}
\label{sec:dataprocessing}
To investigate geospatial decentralization, we acquired validator data, including their locations and stakes, from five leading blockchains: Aptos, Avalanche, Ethereum, Solana, and Sui, as detailed in Table~\ref{tab:data_collection}. This endeavor is nontrivial, as, to our knowledge, we are among the first to empirically compile such comprehensive data.

We collected data primarily through APIs from public explorers~\cite{solanabeach2024, aptoslabs2024, avascan2024}, with Sui’s data shared privately upon request\footnote{Data for Sui was provided through personal communication with Alberto Sonnino, Mysten Labs.}. In cases where explicit location data was unavailable, we estimated validator geolocations using IP addresses~\cite{ipinfo2024}. While we assume the accuracy of these sources, both IP-based geolocation and public explorer data can be imprecise due to VPN usage or outdated information. However, we assume that validators have no strong incentive to mask their locations in the current blockchains.

\begin{table}[ht]
\centering
\caption{Validator data collection and pre-processing}
\label{tab:data_collection}
\begin{tabular}{|l|l|c|c|c|}
\hline
\textbf{Blockchain} & \textbf{\begin{tabular}[c]{@{}l@{}}Data\\ collection\end{tabular}} & \textbf{\begin{tabular}[c]{@{}l@{}}IP-based \\ geolocation\end{tabular}} & \textbf{\begin{tabular}[c]{@{}l@{}}Validator\\ count\end{tabular}} & \textbf{\begin{tabular}[c]{@{}l@{}}Validator\\count after\\ pre-processing\end{tabular}} \\ \hline \hline
Ethereum & \begin{tabular}[c]{@{}l@{}}Beacon node \\ subnets\end{tabular} & false & 10,803 & 1,046 \\ \hline
\begin{tabular}[c]{@{}l@{}}Ethereum \\ nodes\end{tabular} & \begin{tabular}[c]{@{}l@{}}Web \\ scraping~\cite{bitfly2024}\end{tabular}   & true & 5,402 & 875 \\ \hline
Solana & API~\cite{solanabeach2024} & true & 2,310 & 118 \\ \hline
Aptos & API~\cite{aptoslabs2024} & \centering false & 186 & 46 \\ \hline
Sui & \begin{tabular}[c]{@{}l@{}}Shared\\ upon request\textsuperscript{2}\end{tabular} & false & 48 & 47 \\ \hline
Avalanche & API~\cite{avascan2024} & true & 1,468 & 99 \\ \hline
\end{tabular}
\end{table}

\subsection{Ethereum Data Collection Methodology}
We initially used the Ether nodes API to gather validator geospatial data~\cite{bitfly2024}, marked as Ethereum nodes in Table~\ref{tab:data_collection}, but it was not suitable because it could not differentiate between full nodes and validators.
To improve data accuracy, we monitored beacon nodes, which coordinate Ethereum’s consensus. Validators subscribe to short-lived subnets of beacon nodes when assigned as attestation aggregators during an epoch. Tracking these subnet subscriptions across multiple epochs allowed us to estimate the number of validators per beacon node. the validator locations were inferred from IP addresses~\cite{ipinfo2024}.

Assuming rational behavior, we consider all validators to hold 32 ETH, as staking more do not provide additional benefits~\cite{john2024economics}. However, the method has limitations with under-reporting, as we can only record up to 62 validators due to the maximum number of short-lived subnets we can track per beacon node. Additionally, we cannot account for nodes without open P2P ports. Although this method has limitations, we believe that it provides an accurate estimate for Ethereum validator geospatial data at present. This is marked as Ethereum in Table~\ref{tab:data_collection}.

\subsection{Data cleaning and Pre-processing}

During pre-processing, we excluded validators with missing location data. The total stake ignored, along with the number of validators excluded, was 31.57\% (482) for Avalanche, 5.5\% (49) for Aptos, and 0.07\% (931) for Solana.

We focus on the distribution of stake across locations rather than the absolute number of validators, as we analyze global geospatial trends. To enable accurate global geospatial analysis, validators in close proximity, i.e., 20km radius, are merged. By precomputing the distance between all validator pairs using their geospatial coordinates, we identify the validators in close proximity to merge. When merging, the stake weights of two validators are summed, and one of the validators (the one with the lower stake) is then removed from the dataset. This approach helps to mitigate the impact of local variations, such as neighboring data centers. Table~\ref{tab:data_collection} provides a detailed breakdown of validator counts and processing across blockchains. The dataset, including raw validator geolocations, stakes, and scripts for pre-processing (e.g., proximity merging and stake aggregation), is shared to facilitate reproducibility.
Despite inherent limitations, our dataset
\footnote{https://github.com/MSRG/validators-geodata}
offers the most accurate geospatial validator data currently available for Ethereum and other blockchains. In the following section, we focus on quantifying geospatial decentralization.

\section{Quantifying Geospatial Decentralization}

We first define the properties required for a metric to capture geospatial decentralization, then evaluate existing decentralization metrics and propose the Gini of Eigenvector Centrality (GEC) metric. Finally, we apply GEC to the empirical data to quantify geospatial decentralization.
 
\subsection{Properties of a Geospatial Decentralization Metric}
\label{sec:properties-metric}

We seek a metric \( M(\mathcal{V})\) to measure geospatial decentralization for a given blockchain with a validator set \(\mathcal{V}\) at epoch \( t \) that satisfy three key properties:
\begin{enumerate}
    \item \textbf{Quantifiability:}
    \( M(\mathcal{V}) \) should yield a scalar in \(\mathbb{R}\), enabling direct comparisons across blockchains and over time.

    \item \textbf{Inequality sensitivity:}
    \( M(\mathcal{V})\) should decrease if a small subset of validators holds a disproportionately large fraction of total voting power \(\mathcal{R}\) in a confined region, thereby reducing decentralization.

    \item \textbf{Geospatial awareness:}
    \( M(\mathcal{V})\) should account for geospatial dispersion, assigning higher values when validators are distributed across distant regions than when they are clustered.

\end{enumerate}

In summary, an ideal geospatial decentralization metric is a scalar function \( M : \mathcal{V} \mapsto \mathbb{R} \) that is quantifiable, sensitive to inequality, geospatially aware, and system-wide. The following subsections review existing metrics, illustrate their shortcomings, and motivate the introduction of a new metric that satisfies all these properties.

\subsection{Assessment of Existing Decentralization Metrics}
\label{sec:existing-metrics}

Numerous decentralization metrics have been proposed in the literature~\cite{kwon2019impossibility,motepalli2024does,schneider2003decentralization,gencer2018decentralization,lin2021measuring}. In this section, we assess these metrics against the desired properties outlined in Section~\ref{sec:properties-metric}, as summarized in Table~\ref{tab:metrics_comparison}.

\begin{table}[h]
\centering
\caption{Decentralization Metrics vs. Desired Properties}
\label{tab:metrics_comparison}
\begin{tabular}{|l|c|c|c|}
\hline
\textbf{Metric} & \textbf{Quantifiability} & \textbf{\shortstack{\\Inequality\\ Sensitivity}} & \textbf{\shortstack{\\Geospatial\\ Awareness}}  \\ \hline \hline
\shortstack{\\Nakamoto coefficient}  & \cmark & \cmark & \xmark \\ \hline
\shortstack{\\Gini coefficient} & \cmark & \cmark & \xmark \\ \hline
\shortstack{\\Entropy} & \cmark & \cmark & \xmark  \\ \hline
\shortstack{\\KDE} & \xmark & \xmark & \cmark \\ \hline
\shortstack{\\Moran’s I} & \cmark & \xmark & \cmark  \\ \hline
\end{tabular}
\end{table}

\textit{Nakamoto Coefficient} measures the minimum number of validators required to compromise a blockchain's safety or liveness~\cite{nakamoto2008bitcoin,motepalli2024does}. While effective in assessing voting power concentration, it ignores the geospatial distribution. For example, blockchain B with a coefficient of 100 is considered more decentralized than blockchain A with 20. However, if A's validators are globally dispersed and B's are centralized in a single data center, B is geospatially centralized, thus the Nakamoto coefficient fails to capture geospatial decentralization.

\textit{Gini coefficient} quantifies inequality in voting power distribution~\cite{gini1921measurement,motepalli2024does}, and \textit{entropy} measures the diversity in voting power~\cite{sharma2024unpacking,wu2019information}. While both capture disparities in voting power allocation, neither considers validator geography, thereby limiting their effectiveness in evaluating geospatial decentralization.

Furthermore, geospatial metrics such as \textit{Kernel Density Estimation (KDE)} plots the spatial distribution of the voting power~\cite{haining2003spatial}. While it uses geospatial dimension, KDE neither provides a single scalar value for direct comparison nor inherently captures inequalities in voting power distribution.

Additionally, spatial autocorrelation metrics such as \textit{Moran’s I}~\cite{haining2003spatial} measure geospatial correlation in voting power distribution. However, they are insensitive to voting power disparities. For example, a blockchain with a few concentrated clusters of high voting power may be considered equivalent to a geospatially decentralized blockchain, as both lack clear spatial correlation patterns.

None of these metrics simultaneously satisfy the three essential properties of geospatial decentralization outlined in Section~\ref{sec:properties-metric}. To address this gap, we propose the Gini coefficient of eigenvector centrality (GEC), a novel metric specifically designed to meet these criteria, as detailed in the following section.

\subsection{Gini of Eigenvector Centrality (GEC)}

We propose the \textit{Gini of Eigenvector Centrality (GEC)} to analyze the geospatial decentralization of voting power. This metric integrates geospatial proximity and stake-based influence. Eigenvector centrality, widely used in graph theory and foundational to algorithms like PageRank~\cite{ruhnau2000eigenvector, bianchini2005inside}, quantifies a validator's influence based on its stake and proximity to other high stake validators~\cite{bonacich2007some} (see Appendix~\ref{appendix}). Validators closer to others with significant voting power contribute more effectively to reaching consensus quorum \(\mathbb{Q}\).

Each validator \(v_i \in \mathcal{V}\) is modeled as a node in a graph, with \(\Delta_{ij}\) representing the distance between validators \(v_i\) and \(v_j\). The edge weight \(d_{ij}\), derived from a normalized distance matrix, prioritizes proximity:
\begin{equation}
d_{ij} = 1 - \frac{\Delta_{ij}}{\Delta_{\text{max}}}, \quad \Delta_{\text{max}} = \max_{v_i,v_j \in \mathcal{V}} \Delta_{ij}.
\end{equation}

The weighted adjacency matrix \(A\) is defined as:
\begin{equation}
A[i][j] = \rho_i \cdot \rho_j \cdot d_{ij},
\end{equation}
where \(\rho_i\) and \(\rho_j\) denote the voting powers of validators \(v_i\) and \(v_j\), respectively.

Eigenvector centrality scores are computed by solving:
\begin{equation}
A \cdot x = \lambda \cdot x,
\end{equation}
where \(\lambda\) is the principal eigenvalue of \(A\) and \(x\) its corresponding eigenvector. The component \(x[i]\) represents the geospatially weighted centrality score of validator \(v_i\).

To measure inequality in these centrality scores, we compute the Gini coefficient, defining the GEC metric. GEC satisfies key decentralization criteria: quantifiability, sensitivity to geospatial clustering, and incorporation of stake-based voting power. As a holistic metric, GEC is applied to evaluate geospatial decentralization in blockchain systems.

\subsection{Empirical Analysis}

\begin{figure*}[ht]
    \centering
    \includegraphics[width=\textwidth, clip]{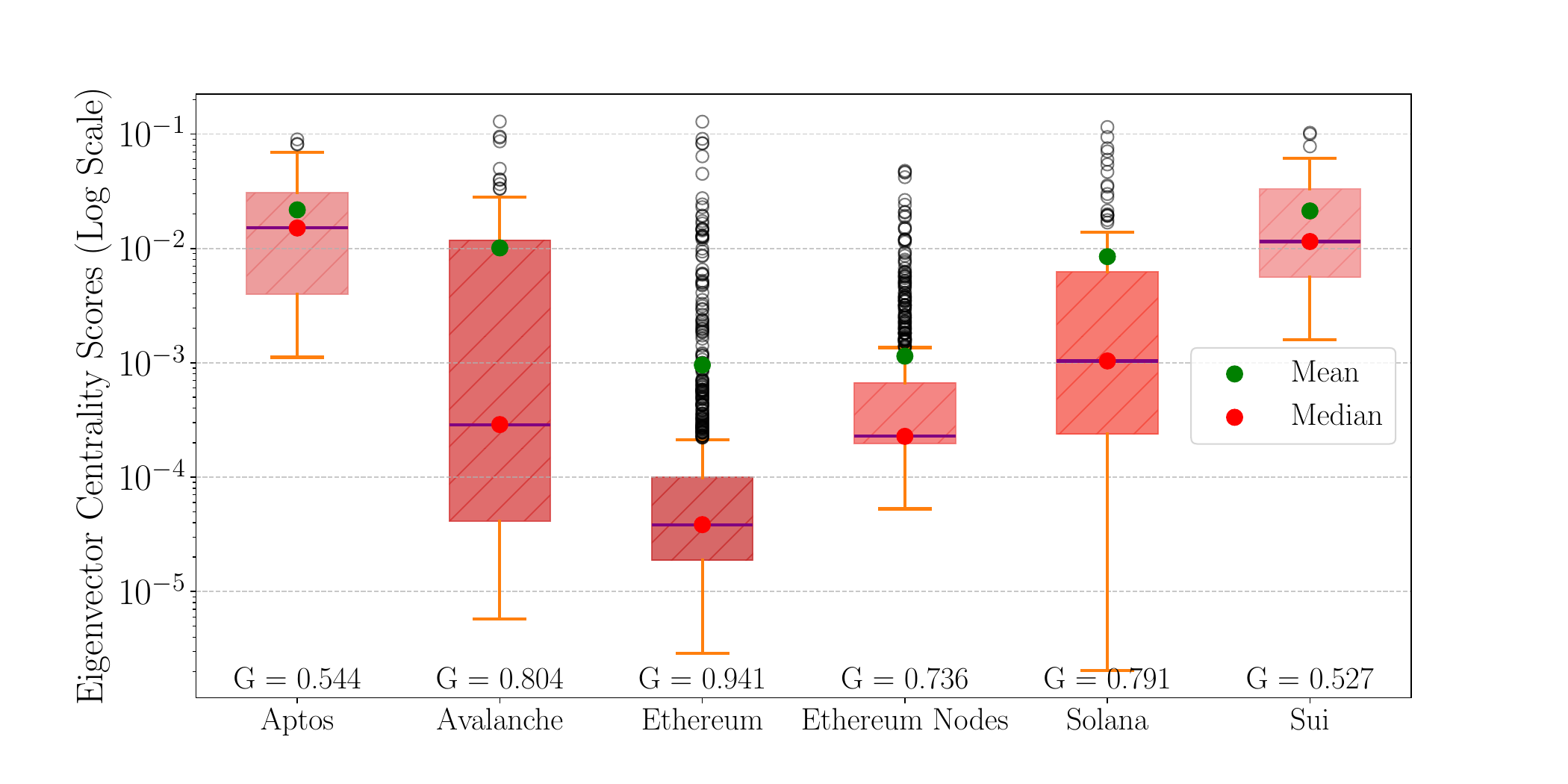} 
    \caption{Boxplots illustrating the distribution of eigenvector centrality measures, showing percentiles, mean, median, and Gini coefficients.}
    \label{fig:centrality_boxplots}

\end{figure*}

We perform empirical analysis on data collected in Section~\ref{sec:dataprocessing}. The eigenvector centrality scores, plotted on a log scale in Figure~\ref{fig:centrality_boxplots}, reveal significant centralization of influence across blockchains. The wide interquartile ranges observed in blockchains such as Avalanche and Solana indicate that a small number of validators exert disproportionate influence compared to their peers, reflecting a lack of uniform distribution.

The distribution of centrality scores is notably skewed, with a clear disparity between the mean and median values. For instance, in Ethereum, the mean of the centrality scores is substantially higher than their median, suggesting that a small group of validators have outsized influence, further confirming the trend toward centralization.

The Gini coefficients of these centrality measures, used to analyze inequality in influence,  which range between 0.527 and 0.941, reinforce this observation. High values like 0.941 in Ethereum and 0.804 in Avalanche indicate severe inequality in influence distribution. Even the lowest observed Gini coefficient of 0.527 suggests centralization, as a significant portion of influence remains concentrated among a limited number of validators.

This analysis demonstrates that, across the evaluated blockchains, influence—measured in terms of proximity and voting power—is not geospatially decentralized. There is a pressing need to enhance geospatial decentralization in blockchains.

\section{Geospatially-aware Proof of Stake (GPoS)}
\label{sec:gpos}
Our empirical analysis reveals a critical shortcoming in blockchains: the lack of geospatial decentralization. This oversight presents significant risks to regulatory and political resilience~\cite{SECEthereum,censorshipEthereum}, compromises robustness against attacks and failures~\cite{amazonSolana}, and undermines equitable participation~\cite{daian2020flash}. Traditional consensus mechanisms promoted centralization by rewarding performance without considering geospatial distribution~\cite{castro1999practical,zhang2024reaching}. This gap underscores the need for solutions that recognize and incorporate geospatial factors into consensus mechanisms.

To address this gap, we propose the \textit{Geospatially-aware Proof of Stake} (GPoS) mechanism to redefine voting power in consensus. Unlike conventional PoS, which relies solely on stake weight, GPoS incorporates both stake weight and the geospatial distribution of validators in its calculation. By integrating geospatial dimensions, GPoS {aims to enhance} decentralization and foster more resilient blockchain systems.

This section begins by quantifying the geospatial distribution of validators using the geospatial diversity index. We then formalize the calculation of voting power within the GPoS mechanism.

\subsection{Geospatial Diversity Index (GDI)}
To quantify the geospatial diversity of validators, we employ the GDI. This index measures a validator's location relative to the locations of all other validators in the blockchain~\cite{motepalli2023analyzing}. Consequently, the GDI is determined by the overall geospatial distribution of the validator set $\mathcal{V}$ rather than solely by an individual validator's location. This relative measure is essential for assessing each validator's contribution to promoting geospatial decentralization within the blockchain.

The calculation of GDI builds upon the existing literature that measures the distance from a given validator to the nearest two-thirds of the validator set~\cite{motepalli2023analyzing}. This approach quantifies relative validator diversity concerning the specified validator set. However, it does not account for the varying voting powers of validators based on their stake. To enhance the applicability of this index, we extend the GDI to incorporate stake weights, recognizing that different validators exert varying levels of influence based on their stake.

\begin{definition}[Geospatial Diversity Index of a Validator]
The GDI of a validator \(v_k\) quantifies its geospatial diversity relative to other validators' stake distributions. Specifically, it measures the minimum distance from \(v_k\) to the closest set of validators whose combined {normalized stake} meets the quorum requirement for consensus. It can be represented as follows:
\begin{equation}
\text{GDI}_k = \min_{V_C \subseteq \mathcal{V} \setminus \{v_k\}} \left\{ \sum_{v_j \in V_C} \Delta_{k, j} \mid \sum_{v_j \in V_C} s_j + s_k \geq \frac{2}{3} \right\}
\end{equation}
\end{definition}

In this equation, \(\Delta_{k,j}\) denotes the distance between validators \(v_k\) and \(v_j\). \(s_j\) indicates the {normalized stake} of validator \(v_j\) {(as shown in Equation~\ref{eq:pos1})} and \(V_C\) represents the subset of nearest validators necessary to form a PoS quorum \(\mathbb{Q}\).

This focus on the PoS quorum $\mathbb{Q}$ is vital as it drives finality in the consensus mechanism, ensuring transaction integrity. Furthermore, the GDI captures geospatial diversity concerning stake distribution within the system. A high GDI indicates that the given validator is geographically distant from others, while a low GDI suggests proximity to other validators. By leveraging the GDI, we can formalize voting power in the GPoS mechanism, enhancing the robustness and fairness of the consensus mechanisms.

\subsection{GPoS Voting Power Formalization}
\label{subsec:votingpowe}
Traditionally, PoS systems only consider the stake weight. Our motivation with GPoS is to incorporate the diversity of validators through the GDI, alongside stake, into the calculation of voting power.

{In GPoS, we first compute an \textit{intermediate influence score}, $\omega_i$, for each validator $v_i$, which is a function of both its stake and its GDI. Specifically, we adopt a \textit{linear combination} to balance the influence of both factors:}
\begin{equation}
\label{eq:gpos_score}
    {\omega_i} = f(s_i, GDI_i) = \lambda \cdot s_i + (1-\lambda) \cdot GDI'_i
\end{equation}

Here, \( 0 \leq \lambda < 1 \) is a tunable weight parameter. {\( s_i \) represents the normalized stake (as calculated with Equation~\ref{eq:pos1})} and \( GDI'_i \) is the \textit{normalized GDI}, where normalization is done as follows:
\begin{equation}
\label{eq:gdi_norm}
\quad GDI'_i = \frac{GDI_i}{\max(GDI)}
\end{equation}
Here, $\max(GDI)$ represents the maximum value of GDI among all validators. {To ensure the total voting power sums to one, consistent with BFT consensus requirements, we normalize these influence scores to calculate the final GPoS voting power, $\rho_i^*$:}
\begin{equation}
\label{eq:gpos_final}
{
    \rho_i^* = \frac{\omega_i}{\sum_{v_j \in \mathcal{V}} \omega_j}
}
\end{equation}

{GPoS is designed to augment, not replace, the foundational security principles of Proof of Stake. The core of PoS security is the principle of capital-at-risk: voting power is directly proportional to economic stake, which can be slashed.} The parameter \( \lambda \) determines the relative weight assigned to stake versus GDI. When $\lambda=1$, GPoS is traditional PoS. {We recommend constraining $\lambda$ to the range $[0.5, 1)$ as a deliberate security design choice. Setting $\lambda < 0.5$ would allow the geospatial factor to contribute more to voting power than stake. This could enable an adversary with a minority of the network's stake ($<1/3$) to amass a majority of voting power ($>1/3$) by optimizing validator locations, thereby breaking the fundamental economic security of the protocol. The $\lambda \ge 0.5$ constraint ensures stake remains the primary determinant of power, more aligned with the security guarantees of PoS.}

{While GPoS inherits the structure of the underlying BFT protocol, we acknowledge that a formal proof of safety and liveness under our modified voting power distribution requires a separate theoretical analysis, which we leave for future work.}

While other combinations, such as exponential formulations, could be considered, we adopt the linear combination for its simplicity and interpretability. The linear model provides an intuitive and flexible framework for incorporating geospatial diversity into voting power calculations.

\subsection{GPoS Implementation}
PoS is widely adopted, and transitioning to GPoS is straightforward. As in most blockchains, the validator set $V = \{v_1, v_2, \dots, v_n\}$ remains fixed during an epoch $t$ and is updated only at the start of the next epoch $t+1$ through a \textit{reconfiguration} mechanism. This mechanism updates both the set of validators and their voting power \( \rho_i \), using staking data $s_i$, slashing criteria, and geospatial coordinates $c_i(x_i, y_i)$, which are assumed to be available on-chain. Since GPoS modifies the computation of \( \rho_i \) by integrating the GDI, our focus is on the reconfiguration mechanism.

In BFT PoS chains, reconfiguration happens at epoch boundaries; in CometBFT, the ABCI app returns \textsc{ValidatorUpdates} in \textsc{EndBlock} with voting powers $\rho_i$. The updated reconfiguration mechanism is as follow (Algorithm~\ref{algo}).
\begin{itemize}
    \item Validators query the blockchain to retrieve updated $s_i$ and $c_i(x_i, y_i)$, reflecting changes during the epoch.
    \item The validator set $V$ is determined deterministically based on protocol criteria, such as a fixed size or minimum stake threshold.
    \item {Each validator calculates the GDI,  an intermediate influence score $\omega_i$, and the final normalized voting power $\rho_i^*$.}
    \item The updated validator set and their voting powers are encoded in the block header of the first block of the new epoch.
\end{itemize}

\begin{algorithm}[H]
\caption{Epoch $t \!\to\! t{+}1$ Reconfiguration under GPoS}
\label{algo}
\begin{algorithmic}[1]
\REQUIRE Validator candidates with stake $s_i$ \& coordinates $c_i(x_i,y_i)$; protocol parameter $\lambda\in[0,1)$
\STATE \textbf{On epoch boundary} $t \to t{+}1$:
\STATE Read on-chain $s_i$ (post-slashing) and $c_i$ updated during epoch $t$
\STATE $V \gets \textsc{DeterministicSelectEligible}( \{s_i,c_i\} )$ \COMMENT{e.g., top-$K$ or min-stake threshold}
\FOR{$v_i \in V$}
    \STATE $GDI_i \gets \textsc{ComputeGDI}(v_i, V)$
\ENDFOR
\STATE {$GDI_{max} \gets \max_{j \in V}(GDI_j)$}
\FOR{$v_i \in V$} 
    \STATE {$GDI'_i \gets GDI_i / GDI_{max}$}
    \STATE {$\omega_i \gets \lambda \cdot s_i + (1-\lambda) \cdot GDI'_i$}
\ENDFOR
\STATE $V' \gets \{(v_i, {\rho_i}) \mid v_i\in V\}$ where {$\rho_i \gets \omega_i /\sum_{v_j \in V} \omega_j$} 
\COMMENT{validator set and voting power for epoch $t{+}1$}
\STATE \textbf{Commit} $V'$ in the header of the \emph{first block of epoch $t{+}1$}
\end{algorithmic}
\end{algorithm}

Every validator \(v_i\) adds their geospatial coordinates \(c_i(x_i, y_i)\) on-chain. Validator locations are already publicly accessible in permissionless networks for peer discovery~\cite{heimbach2024deanonymizing, solanabeach2024, avascan2024, aptoslabs2024}, thus requiring to publish coordinates in GPoS does not introduce additional privacy or security vulnerabilities~\cite{heilman2015eclipse}. The coordinates of the validators are assumed to be correct unless disputed. Similar to optimistic rollups~\cite{motepalli2023sok,kalodner2018arbitrum}, we employ a dispute resolution mechanism.

\begin{itemize}
    \item \textit{Dispute Initiation}: Any validator \(v_j\) can challenge \(c_i\) by submitting a dispute claim with collateral {\(S_j^{\text{dispute}}\) }(e.g., 10\% of their stake). This claim must include external proofs derived from triangulation techniques, oracle services, or Proof-of-Location systems~\cite{sheng2024bft}.
        
    \item The validator set \(V\) interacts with the dispute contract through the underlying consensus mechanism to achieve a quorum. 
    \item \textit{Outcome}: 
        \begin{itemize}
            \item If \(c_i\) is proven invalid: \(v_i\)'s stake {\(S_i\)} is slashed, with 20\% of the slashed amount awarded to \(v_j\).
            \item If \(c_i\) is valid: \(v_j\)'s collateral {\(S_j^{\text{dispute}}\)} is burned for initiating a false dispute.
        \end{itemize}
\end{itemize}

{This mechanism creates a strong economic disincentive against location spoofing. While challengers bear the cost of gathering external proof, the high reward for a successful challenge (a significant portion of the slashed stake) motivates validators to police one another. Conversely, the risk of losing substantial staked collateral in a failed challenge disincentivizes frivolous disputes, creating a balance of economic incentives.} Empirical data from optimistic rollup deployments indicate that dispute resolution requires only a single on-chain transaction of approximately 25{,}000--300{,}000 gas (\(\approx \$0.01\)--\$15 at prevailing prices)~\cite{liu2024fairrelay}, and disputes occur in fewer than 0.01\% of transactions~\cite{lee2025looking}, making the mechanism economically and operationally negligible.


\subsubsection{{Security Considerations: Sybil Attack Resistance}}
\label{sec:sybil_resistance}
{A critical security consideration is a Sybil attack where an adversary creates numerous low-stake validators with spoofed, geographically diverse locations to unfairly gain voting power. GPoS mitigates this threat through its stake-weighting mechanism. The final voting power $\rho_i^*$ is a function of both stake and GDI, governed by $\lambda$, which we recommend setting at $\lambda \geq 0.5$. This ensures that stake remains the dominant factor in consensus.}

{Let an adversary control a total normalized stake of $s_{adv}$, distributed across any number of Sybil validators. Their collective influence score, $\Omega_{adv} = \sum \omega_j$, consists of a stake component, $\lambda s_{adv}$, and a geospatial component, $(1-\lambda)\sum GDI'_{j}$. While an adversary can maximize the geospatial term by spoofing ideal locations, its overall weight is capped by $(1-\lambda)$. Since we set $\lambda \ge 0.5$, an adversary's voting power is always fundamentally constrained by its capital stake ($s_{adv}$), preventing it from gaining disproportionate control. An attack with near-zero-stake Sybils is thus ineffective, as its influence remains negligible.}

\subsection{Computational Complexity of GPoS}
Each validator recomputes its geospatial weight once per epoch via two steps. i) Pairwise distance matrix computation among $n$ active validators, with $O(n^2)$ complexity. ii) Per‐validator GDI calculation, which involves sorting distances and selecting a quorum, at $O(n\log n)$ complexity. Thus, the combined worst‐case time complexity is $O(n^2 \log n)$.

Despite the quadratic term, real‐world performance remains practical due to:
\begin{itemize}
  \item Symmetry: $\Delta_{ij}=\Delta_{ji}$ halves the required distance computations.  
  \item Parallelization: Distance calculations can be distributed across cores.  
  \item Caching \& Incremental Updates: Validator membership changes infrequently; we persist the previous distance matrix and recompute only for joining or departing validators.
\end{itemize}
We ran our experiments up to 10,000 validators, exceeding typical validator-set sizes (hundreds to low thousands) observed across major blockchains (see Table~\ref{tab:data_collection}). On commodity hardware, our optimized implementation computes GDI for $n=10{,}000$ in under 60s~\footnote{\url{https://github.com/GeoDecConsensus/geo-analysis/blob/main/data/gdi_complexity_report.md}}. Since epochs span $\approx$24 h, a sub‐minute offline computation imposes negligible overhead. Moreover, incremental updates further reduce both CPU and memory costs in practice. Together, these optimizations ensure that, although the theoretical complexity is $O(n^2 \log n)$, GPoS remains highly scalable for large validator sets.

\subsection{Consequences of GPoS}

GPoS implementation introduces several consequences that enhance the resilience of blockchains. We examine two major effects.

\subsubsection{Proposer Selection}

In most consensus mechanisms, the block proposer is selected based on voting power. The probability \( P_i \) of validator \( v_i \) being chosen as a proposer is:

\begin{equation}
    P_i = \Phi \cdot \rho_i
\end{equation}where \( \Phi \) represents a protocol-defined randomness factor, and \( \rho_i \) is the voting power of validator \( v_i \).

Under GPoS, voting power incorporates geospatial diversity, encouraging proposer selection from diverse locations. This reduces the likelihood of latency-driven front-running, enhancing fairness for end-users. It also broadens opportunities for validators across regions to participate in MEV capture.

\subsubsection{Reward Distribution}
In PoS, validator \( v_i \) receives a reward \( r_i \) for providing economic security through stake. This results in compounding benefits for high-stake validators and contributes to geospatial centralization. In GPoS, security is defined not only by stake, but also by geospatial diversity, as reflected in voting power \( \rho_i^* \). This adjustment reduces the compounding effects of regional concentration and fosters long-term geospatial decentralization. In the following section, we empirically evaluate how GPoS affects geospatial decentralization relative to PoS.

\subsubsection{Strategic Incentives and Validator Behavior}
\label{sec:incentives}
{By design, GPoS alters validator economic incentives to favor geospatial diversity.  This introduces potential strategic behaviors, which are mitigated by the protocol's design. Malicious strategies such as \textbf{location spoofing} is disincentivized by the dispute mechanism (Section 5.3), where the high economic penalty of slashing deters fraud. Furthermore, as detailed in Section 5.3.1, the impact of \textbf{Sybil attacks} is constrained by the stake-weighting parameter ($\lambda \ge 0.5$), which ensures an adversary's voting power remains coupled to their capital at risk.}

{Conversely, strategic validator relocation to underserviced regions to gain a higher GDI score is not a form of gaming but rather an intended, desirable outcome of the GPoS mechanism, as it promotes greater decentralization. While our mitigations address immediate attack vectors, a formal game-theoretic analysis of the long-term validator behaviors under GPoS is an important area for future research.}
\begin{figure*}[]
    \centering
    \includegraphics[width=\textwidth, trim=10 0 2 10, clip]{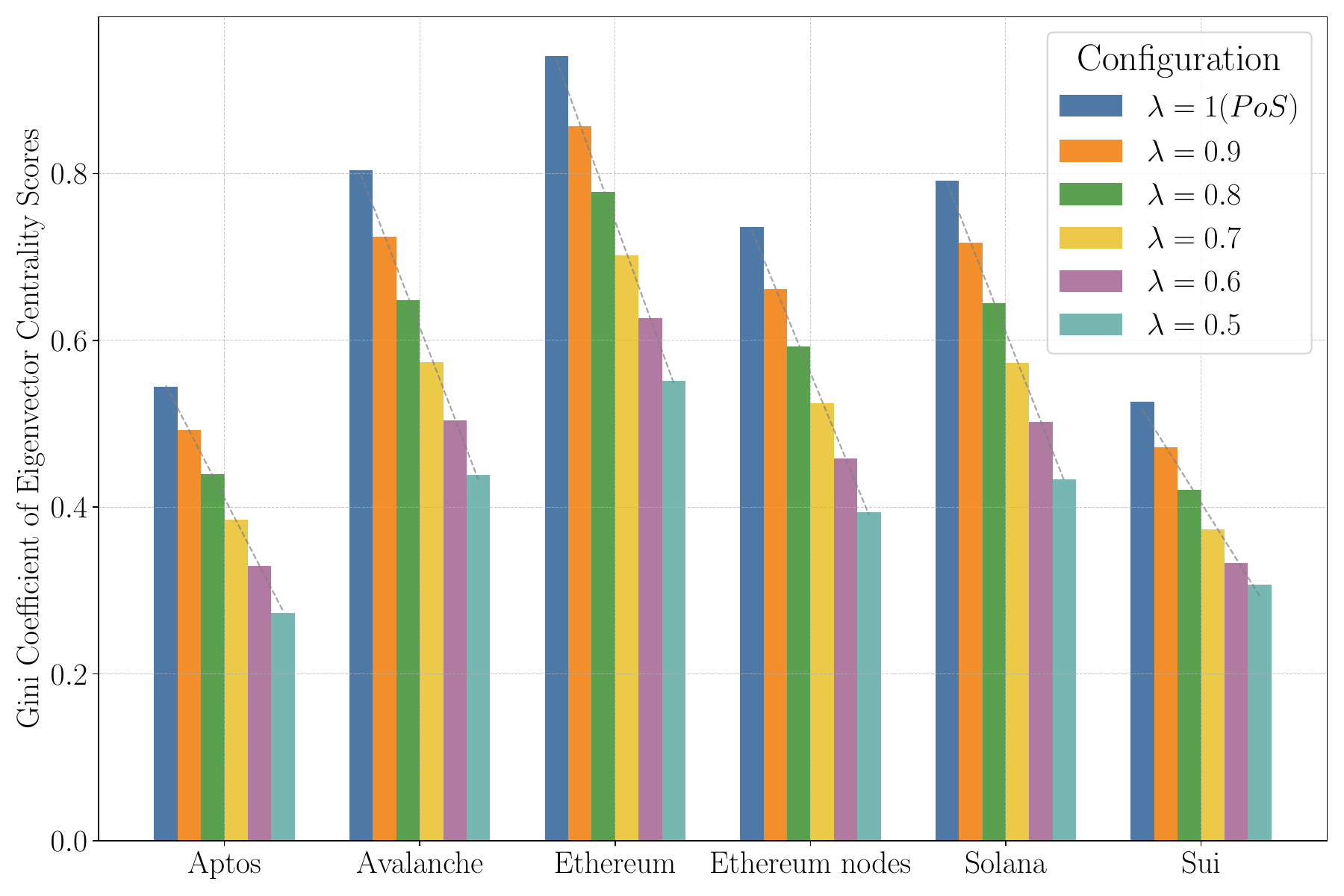} 
    \caption{Gini coefficients for eigenvector centrality scores across blockchains, with varying \( \lambda \) values.}
    \label{fig:centrality_gpos}
\end{figure*}

\subsection{Empirical Evaluation}
\label{sec:empirical}
To evaluate the effectiveness of GPoS in improving geospatial decentralization compared to traditional PoS, we conducted empirical analysis. 

\subsubsection{GEC}
We analyze GEC to quantify geospatial decentralization. Figure~\ref{fig:centrality_gpos} presents the Gini coefficient for the eigenvector centrality scores across blockchains. The parameter \(\lambda\) was varied from 0.5 to 0.9 in increments of 0.1 to capture its effect.

The results show that transitioning from traditional PoS (\(\lambda = 1\)) to GPoS (\(\lambda = 0.5\)) consistently reduces the Gini coefficient across all blockchains, indicating increased geospatial decentralization. On average, the Gini coefficient decreased by 45\%, with individual reductions ranging from 41.38\% to 49.72\%. These results demonstrate that GPoS effectively redistributes the influence of the validator, mitigating the typical PoS concentration.

\begin{figure*}[]
    \centering
    \includegraphics[width=\textwidth, trim=5 10 2 5, clip]{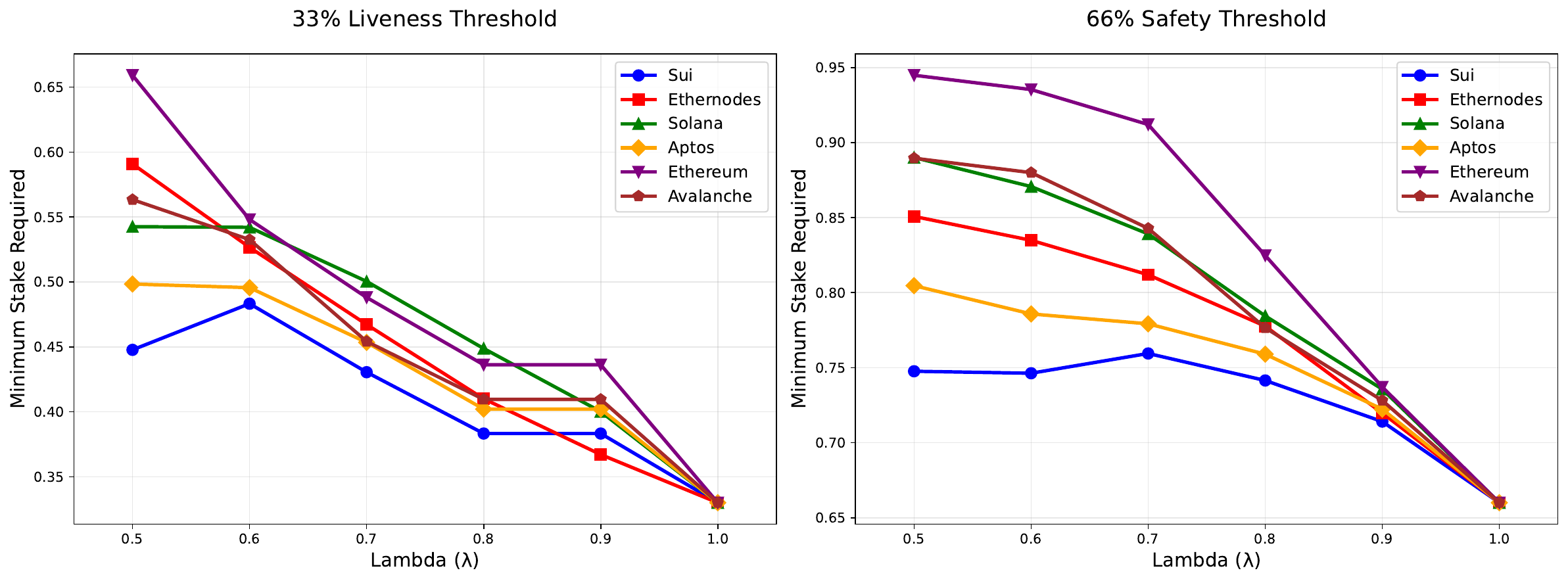} 
    \caption{Minimum stake required for GPoS is higher than PoS (\( \lambda =1 \)), across blockchains.}
    \label{fig:stake_required}
\end{figure*}

\subsubsection{Stake weight}
We quantify the minimum stake required to violate liveness (33\%) and safety (66\%) thresholds under current PoS and GPoS in Figure~\ref{fig:stake_required}. For all evaluated blockchains, as $\lambda$ decreases (increasing the weight of GDI), the minimum adversarial stake required to reach these thresholds increases, assuming an adversary can optimally distribute stake across locations. {Our analysis in Figure~\ref{fig:stake_required} indicates that security against coalition attacks under GPoS requires an equal or greater adversarial stake than} traditional PoS, given the current validator distributions.

\section{Experimental Evaluation}
\label{sec:experimental}
Our objective is to assess how the GPoS mechanism impacts the performance of consensus protocols, measured in throughput and latency. We emulate two prominent BFT consensus mechanisms: HotStuff~\cite{yin2019hotstuff} and CometBFT (formerly known as Tendermint~\cite{buchman2016tendermint}). Both mechanisms are leader-based BFT protocols that provide absolute instantaneous finality.

HotStuff assumes a fully connected network topology and employs a broadcast protocol for communication among validators, with the leader coordinating communications. In contrast, CometBFT utilizes a gossip protocol, requiring validators to communicate with only a subset of their connected peers.

\subsection{System Configurations}

Our experimental setup consists of 64 virtual machines, each with 2 vCPUs, 20 GB disk, and 7.5 GB RAM. To emulate network latencies between validators, we use \textit{netem}~\cite{hemminger2005network}, emulating locations based on pairwise latency data gathered from 250 \textit{servers} across key global locations~\cite{wonderproxy}.

\subsection{Setup}
For our analysis, we first add the validators' voting power to the nearest available server location. After this process, we have 42 locations for Aptos and 40 for Sui. Other blockchains had more than 64 server locations, so we merged them based on proximity until we reached 64 validators. The maximum merging distance was 94 km for Avalanche, 640 km for Ethereum, 660 km for Ethereum nodes, and 192 km for Solana.

\begin{figure*}[htbp]
    \centering
    \includegraphics[width=\textwidth]{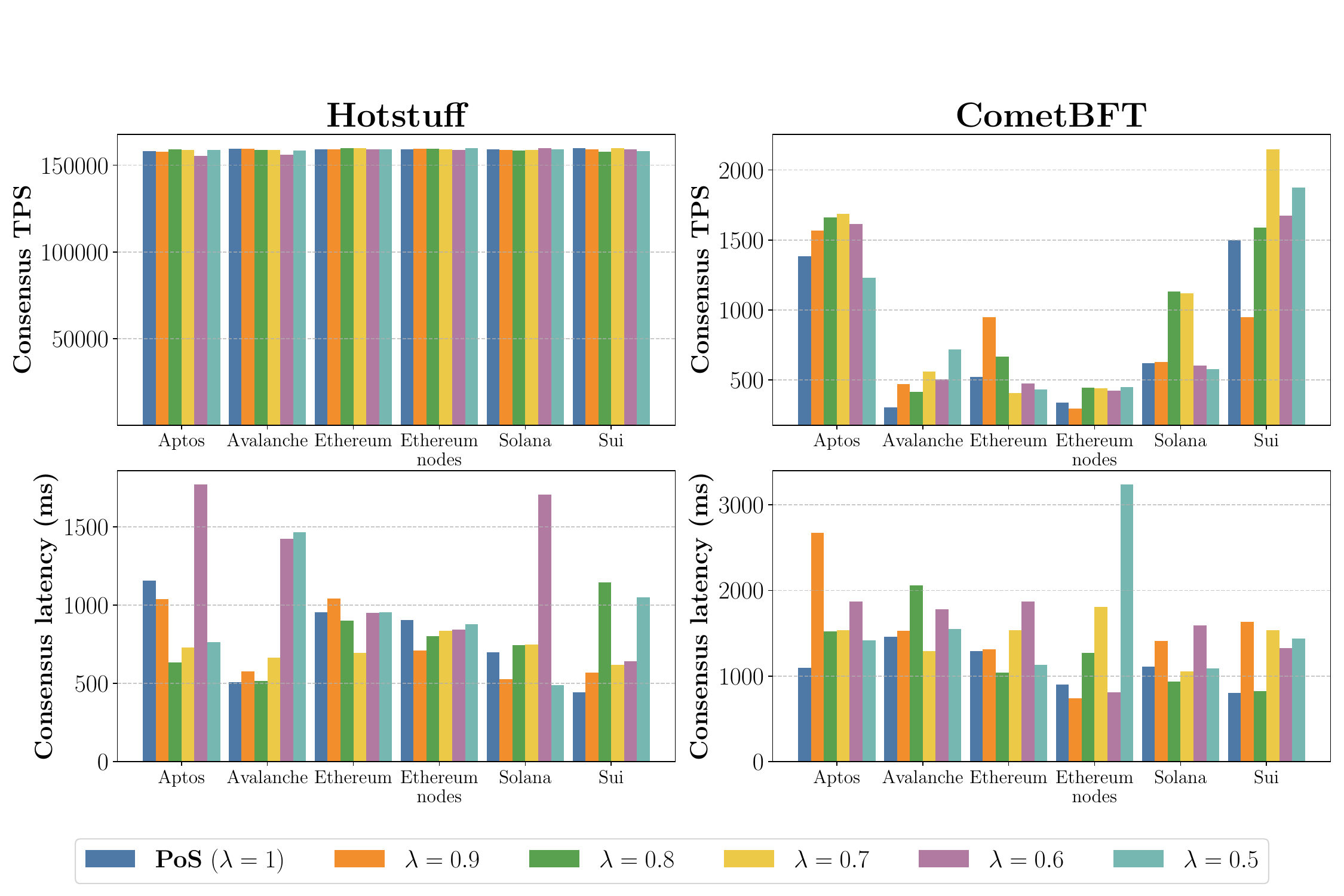}
    \caption{Consensus TPS and Latency Analysis for HotStuff and CometBFT.}
    \label{fig:combined_analysis}
\end{figure*}

For our experiments, we pre-set the latency based on server locations to emulate a wide-area network. All experiments have a fixed message size of 128 bytes, with a consistent batch size and input rate. Clients operated on every server, sending transactions at the same rate. We conducted multiple runs (at least five) for each configuration over a period of 100 seconds to ensure accuracy.

\subsection{Performance Metrics}
We measure the maximum consensus throughput in transactions per second (TPS) and the minimum latency in milliseconds (ms) in the runs, as shown in Figure \ref{fig:combined_analysis}. These experiments reveal distinct characteristics between HotStuff and CometBFT under varying values of $\lambda$. HotStuff demonstrates a consistent TPS of 160,000 in all configurations, indicating stability as geospatial diversity increases. In contrast, CometBFT exhibits greater sensitivity because gossip protocols require multiple rounds of communication, unlike the single round used in broadcast protocols. This sensitivity is further influenced by the specific distribution topology of the validator.

Different values of $\lambda$ reveal varying impacts on latency across consensus mechanisms. HotStuff maintains consistent latency across most blockchain networks, indicating that GPoS can be effectively applied without significant performance degradation, thereby supporting enhanced decentralization. In contrast, CometBFT exhibits greater latency variability, particularly in networks such as Aptos and Ethereum nodes. This sensitivity suggests a trade-off between geospatial decentralization and latency performance, necessitating careful tuning of $\lambda$ and optimization of the consensus mechanism to effectively balance these trade-offs.

\section{Related Work}
\subsection{Geospatially-Aware Consensus Protocols}

Several consensus protocols leverage node geolocation to enhance decentralization or performance. \textit{SENATE}~\cite{jiang2020senate} is a permissionless BFT algorithm designed for wireless IoT networks. It partitions nodes into geographic districts based on wireless network coordinates, electing one representative per district for consensus, thereby ensuring regional fairness and Sybil resistance. Other IoT-oriented variants, such as G-PBFT~\cite{lao2020g} and LH-Raft hierarchical approaches~\cite{guo2022hierarchical}, similarly utilize geographic clustering but in permissioned environments, not permissionless blockchains. \textit{GeoBFT}~\cite{gupta2020resilientdb}, a component of ResilientDB, extends PBFT to geographically distributed deployments by clustering replicas regionally and minimizing cross-region communication to enhance throughput. Unlike SENATE's fixed-per-region delegation or GeoBFT's explicit clustering mechanism, GPoS retains the existing BFT/PoS consensus protocol structure and instead dynamically reweights validators' voting power based on geographic diversity. By continuously adjusting stake weightings, GPoS promotes a geographically balanced validator set without introducing new consensus primitives.

Recent BFT protocols (Mahi-Mahi~\cite{jovanovic2024mahi}, Raptr~\cite{tonkikh2025raptr}) aim to maximize performance (throughput and responsiveness) over wide-area networks. These are fundamentally \emph{performance-centric} and none of them address the censorship or regulatory risks arising from the colocation of the validators that GPoS targets.

Recent studies on Layer 2 (L2) blockchains highlight latency racing in transaction sequencing, focusing on centralized sequencers~\cite{mamageishvili2023shared,mamageishvili2023buying}. In contrast, we emphasize decentralized validator sets.

\subsection{Decentralized Proof-of-Location Systems}
PoL schemes verifiably link blockchain participants to real-world coordinates. Early designs often depended on semi-trusted infrastructure or dedicated hardware, but recent work adopts permissionless, cryptographic approaches~\cite{brito2025decentralized}. For example, Helium’s “Proof-of-Coverage” uses radio beacons among wireless hotspots to validate local coverage before admitting them to its BFT consensus group~\cite{haleem2018decentralized}; FOAM beacons conduct ultrasound or RF handshakes via specialized devices~\cite{cedeno2022geospatial}; and BFT-PoLoc embeds calibrated network-delay triangulation directly into a BFT protocol~\cite{sheng2024bft}. Recent zero-knowledge PoL~\cite{wu2020blockchain} enables privacy-preserving location proofs using zk-SNARKs, while \textit{VerLoc}~\cite{kohls2022verloc} provides verifiable localization without trusted landmarks.

GPoS treats location as an off-chain oracle rather than a core consensus primitive. We assume validators’ positions are established externally (e.g., via IP geolocation or existing PoL services) and simply reweight stake to promote geospatial decentralization. This requires no new hardware, yet can leverage PoL frameworks to audit location authenticity while remaining fully compatible with standard PoS protocols.

\subsection{Geolocation and Decentralization in Blockchain Networks}
Although decentralization has been extensively studied, its geospatial aspect remains underexplored~\cite{kwon2019impossibility,lin2021measuring,bahrani2024centralization}. Empirical work shows that major networks are regionally clustered, i.e., Bitcoin and Ethereum nodes mining power concentrate in a handful of countries, exposing them to correlated outages and regulatory capture~\cite{gencer2018decentralization}. In Section~\ref{sec:dataprocessing}, we extend these analyses with new data from five PoS chains and refined proximity metrics.

Our earlier work introduced the Geospatial Decentralization Index (GDI) and employed network‐wide latency emulation to show how distant validators repeatedly miss strict timeout thresholds—often resulting in slashing—and proposed adaptive timeout mechanisms to mitigate this risk~\cite{motepalli2023analyzing}. In contrast, GPoS tackles geospatial bias at its source by integrating location into the voting‐power calculation itself, rebalancing influence across regions without altering consensus timing or {compromising PoS stake requirements, validated on tested blockchain configurations}.

\subsection{GPoS Compared to Prior Work}
Prior work primarily (i) verifies node locations, (ii) modifies consensus to be geo-aware, or (iii) optimizes BFT protocols for WAN performance. In contrast, GPoS directly integrates geospatial diversity into stake-weighting for permissionless blockchains without altering core consensus mechanisms or weakening Sybil resistance. {Empirical evidence suggests that GPoS introduces minimal overhead} while remaining composable with complementary approaches such as PoL verification and latency-based adaptations.

\section{Conclusions}
\label{sec:conclusions}
This paper presented an empirical analysis of geospatial decentralization in blockchains and introduced the Geospatially-aware Proof of Stake (GPoS) mechanism, which incorporates geospatial diversity into stake-based voting power. Our {empirical} findings indicate significant improvements in decentralization, while {our simulations indicate} minimal performance overhead in the tested BFT protocols.

While our empirical analysis suggests that GPoS improves security properties in the tested configurations, these results constitute strong evidence, not a mathematical proof. We have not provided formal proofs that BFT safety and liveness guarantees hold under the modified voting power distribution. A comprehensive, formal security analysis of GPoS is an important direction for future work.

GPoS offers flexibility through the tunable parameter \(\lambda\), balancing stake and geospatial diversity. Lower \(\lambda\) values prioritize geospatial diversity, while higher values (\(\lambda \approx 0.9\)) retain stake dominance with some geospatial diversity. This adaptability permits blockchains to adjust their decentralization strategies as needed. Further customization is possible via alternative weighting schemes, such as exponential and dynamic models, tailored to validator distributions. 

For PoS blockchains with instant absolute finality, GPoS integrates cleanly into existing reconfiguration steps and incurs negligible computational and network overhead, as demonstrated by empirical results on throughput and latency. Its design is compatible with current production protocols and can be adopted by major PoS chains, including Aptos, Celestia, Cosmos, Polygon, Sei, and Sui, without requiring disruptive consensus changes. With accurate location attestation, our mechanism advances practical improvements in geospatial decentralization and resilience.

Future work will focus on enhancing location accuracy with methodologies like IP traceback~\cite{savage2000practical}, topology-based latency estimation~\cite{gueye2004constraint,katz2006towards}, and VPN detection~\cite{zain2019vpn,guo2020deep}, thereby improving the reliability and security of GPoS-based blockchains.

\begin{acks}
This work was supported in part by NSERC and the PBS Foundation.
\end{acks}

\bibliographystyle{ACM-Reference-Format}
\bibliography{references}

\appendix
\section{Comparison of Centrality Metrics for Geospatial Decentralization}
\label{appendix}
We analyze centrality metrics on the stake–proximity weighted graph
$A[i,j] = \rho_i \rho_j d_{ij}$, where $\rho$ is stake and $d_{ij}$ is normalized proximity. We evaluate three desiderata:
(i) stake sensitivity, (ii) spatial awareness, and (iii) recursive influence relevant to quorum formation.

\textbf{Degree centrality} ($\sum_j A[i,j]$) measures local “strength”; it misses multi-hop influence and can overweight dense clusters.

\textbf{Closeness centrality} measures inverse average \emph{weighted} shortest-path distance; it captures geometric dispersion but ignores stake and treats all nodes equally.

\textbf{Betweenness centrality} ranks nodes by frequency on \emph{weighted} shortest paths; it highlights communication bottlenecks, not voting power or regional diversity.

\textbf{Eigenvector centrality (EC)} quantifies recursive influence: validators close to other influential (high-stake, proximate) validators receive higher scores, directly modeling effects relevant to reaching quorum.

\begin{table}[h!]
\centering
\begin{tabular}{lccc}
\toprule
\textbf{Metric} & \textbf{Stake sensitive} & \textbf{Spatially aware} & \textbf{Recursive influence} \\
\midrule
Degree         & Partial (local) & Local      & No \\
Closeness      & No              & Global     & No \\
Betweenness    & No              & Path-based & No \\
Eigenvector    & Yes             & Yes        & Yes \\
\bottomrule
\end{tabular}
\end{table}

\noindent
Eigenvector centrality uniquely satisfies stake sensitivity, spatial awareness, and recursive influence. The Gini coefficient of eigenvector centrality (GEC) therefore provides a robust, interpretable scalar for quantifying geospatial decentralization.

\section{Empirical Analysis using Geospatial Data}
This chapter analyzes geospatial decentralization across blockchains, focusing on the distribution of voting power within consensus mechanisms. Current decentralization metrics~\cite{kwon2019impossibility,motepalli2023analyzing,schneider2003decentralization}, such as validator set cardinality, the Nakamoto coefficient~\cite{balajidecentralization}, and entropy measures, fail to consider the geospatial dimension. Therefore, we design novel measures to address this gap. By examining stake, the proxy for voting power in PoS blockchains, we assess its geospatial distribution within the collected data.
We already studied GEC in this thesis, here we present the alternatives we considered. 

\subsection{KDE Plots for Blockchains}
\label{appendix:kde}

\begin{figure}[htbp]
    \centering
    \includegraphics[width=\textwidth, trim=30 108 48 128, clip]{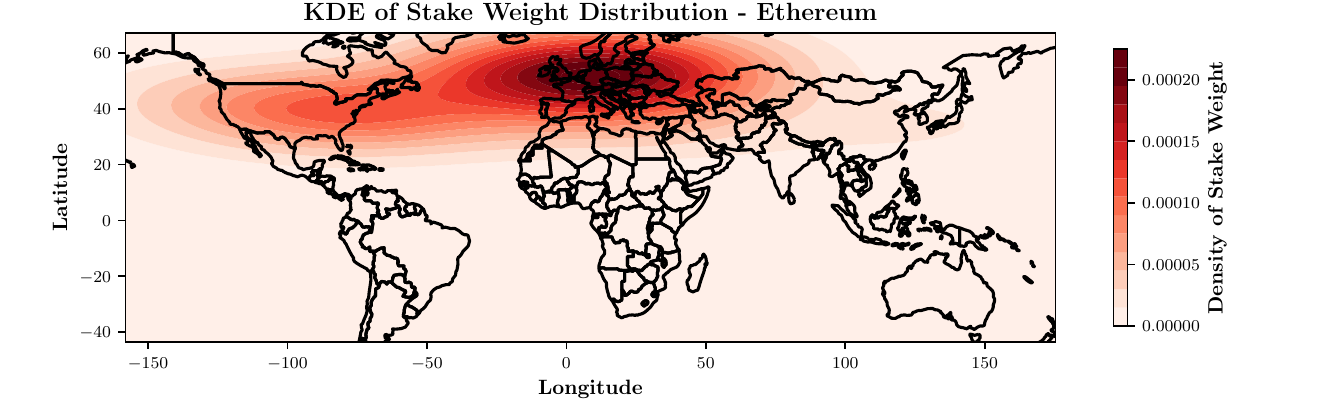}  
    \caption{KDE plot showing Ethereum's stake distribution, with a notable concentration in Europe and North America, indicating potential geospatial centralization.}
    \label{fig:kde_plot}
\end{figure}

\begin{figure}[htbp]
    \centering
    \includegraphics[width=\textwidth, trim=30 108 48 120, clip]{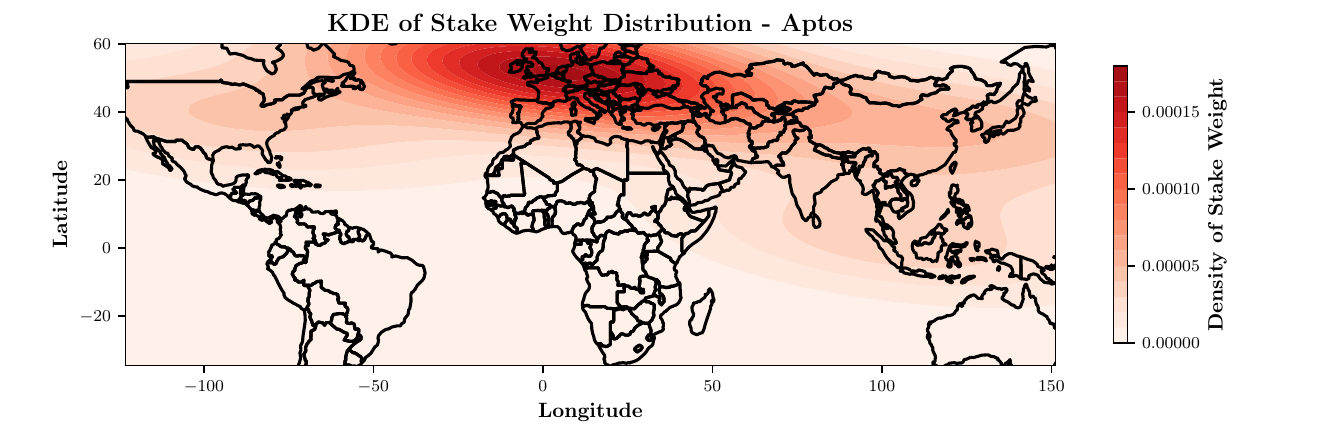}
    \caption{KDE plot showing Aptos stake distribution.}
    \label{fig:aptos_kde_plot}
\end{figure}

\begin{figure}[htbp]
    \centering
    \includegraphics[width=\textwidth, trim=30 108 48 120, clip]{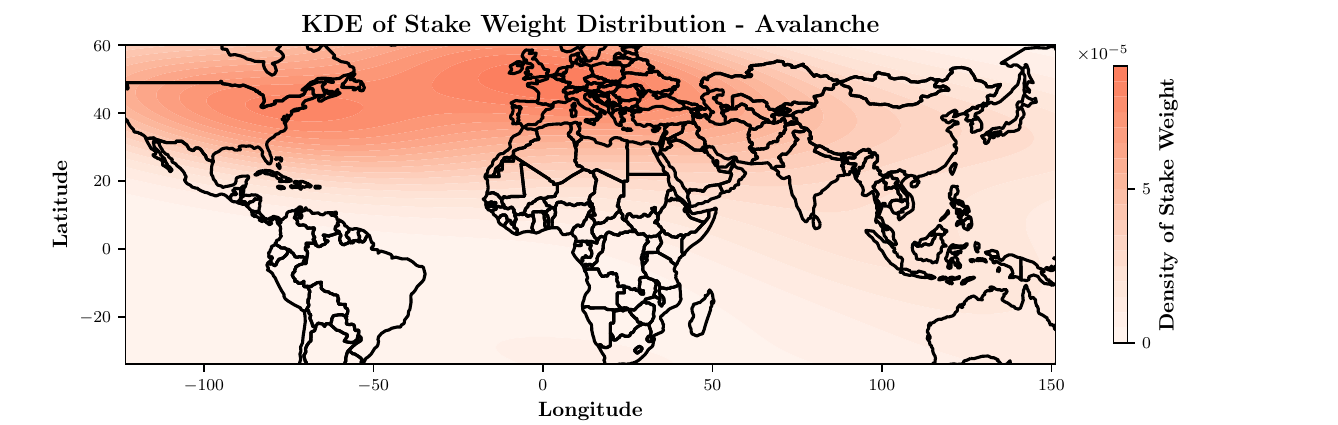}
    \caption{KDE plot showing Avalanche stake distribution.}
    \label{fig:avalanche_kde_plot}
\end{figure}

\begin{figure}[htbp]
    \centering
    \includegraphics[width=\textwidth, trim=30 108 48 120, clip]{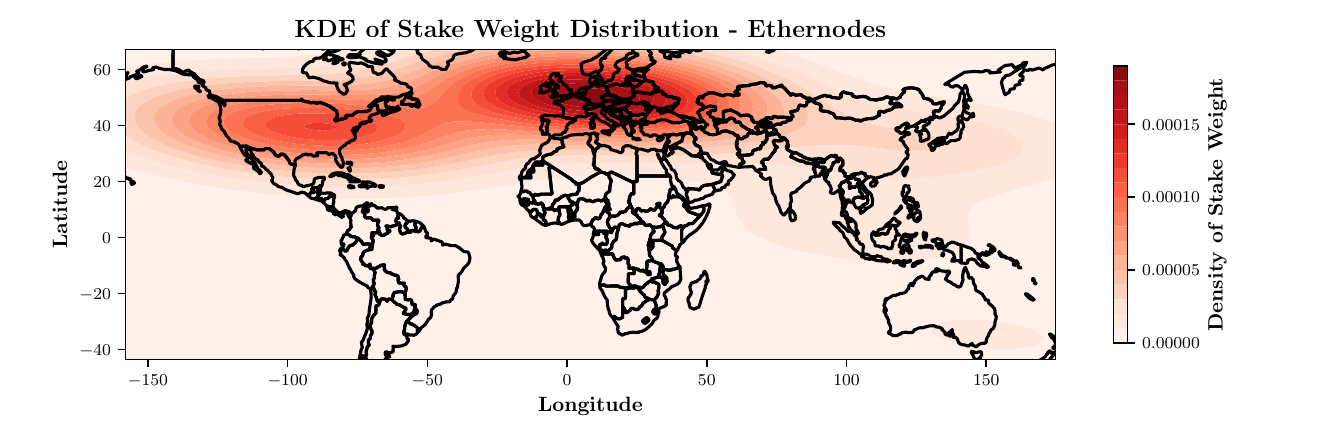}
    \caption{KDE plot showing Ethereum Nodes stake distribution.}
   \label{fig:ethernode_kde_plot}
\end{figure}
\begin{figure}[htbp]
    \centering
    \includegraphics[width=\textwidth, trim=30 100 48 120, clip]{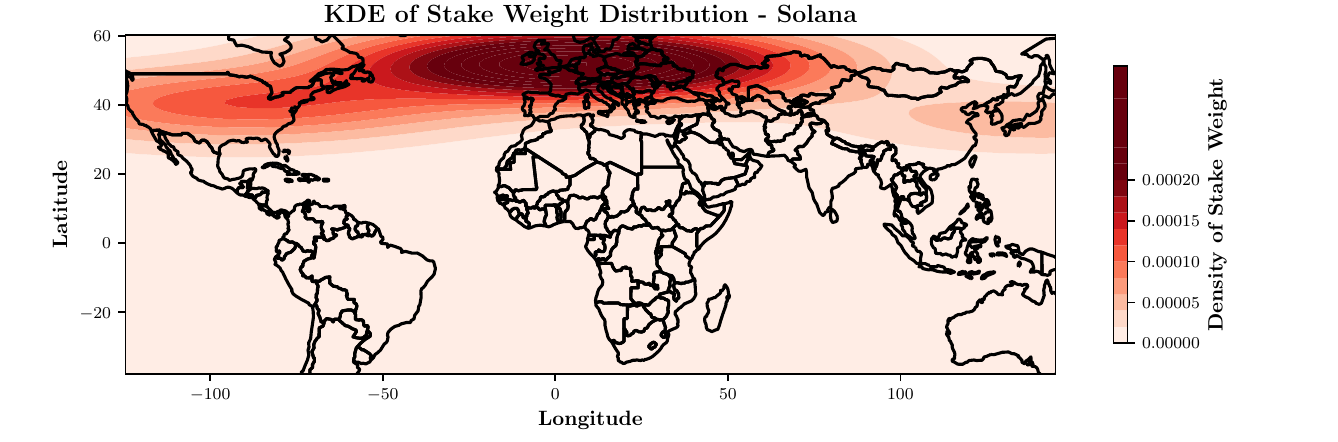}
    \caption{KDE plot showing Solana stake distribution.}
    \label{fig:solana_kde_plot}
\end{figure}

\begin{figure}[htbp]
    \centering
    \includegraphics[width=\textwidth, trim=30 108 48 120, clip]{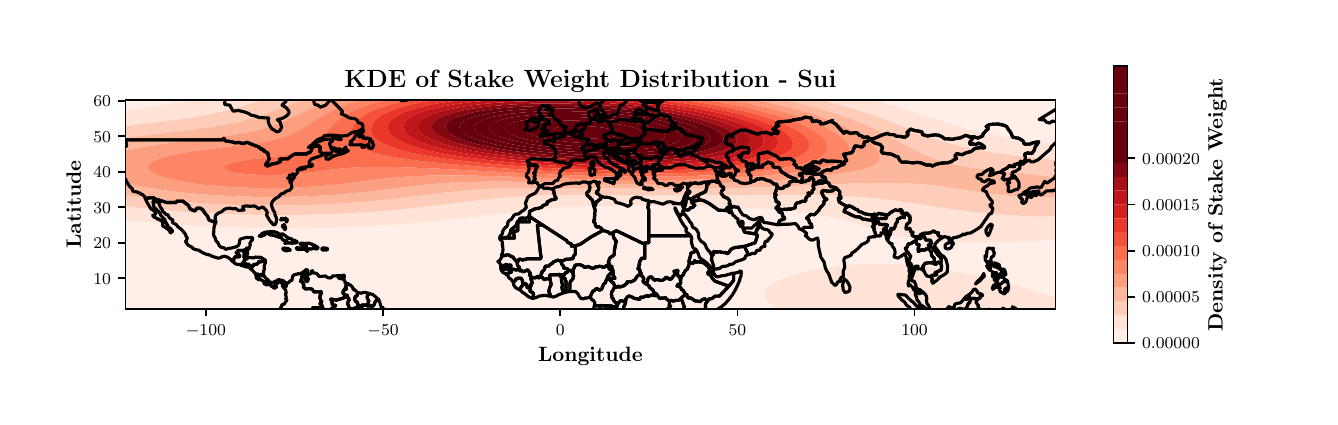}
    \caption{KDE plot showing Sui stake distribution.}
    \label{fig:sui_kde_plot}
\end{figure}

This section presents the Kernel Density Estimation (KDE) plots for various blockchains, illustrating the geospatial distribution of stake weights. Each figure highlights the geographic concentrations of validator influence, providing insights into potential centralization risks.

KDE is a statistical technique used to visualize the distribution of stake weights across geographical regions~\cite{haining2003spatial}. This non-parametric method estimates the probability density function, illustrating areas of stake concentration by employing Gaussian kernels for smoothing~\cite{silverman2018density}.

KDE is crucial for visualizing the geospatial decentralization of blockchains. This method reveals hotspots of stake weights, highlighting potential centralization risks. KDE of stake distribution for Ethereum is illustrated in Fig.~\ref{fig:kde_plot}, indicating significant concentrations in Europe and North America. We observe similar patterns in other blockchains, as shown below. 

\subsection{Gini Coefficient by Country}
The KDE plots presented earlier illustrate significant concentration of stake within select geospatial regions. To quantify these observations, we contextualize the data at the country level. This classification is supported by literature~\cite{de2018blockchain,hinkes2020limits,narayanan2016bitcoin}, which emphasizes the impact of regulatory boundaries on blockchain systems.

\begin{figure}[htbp]
  \centering
  \includegraphics[width=\linewidth]{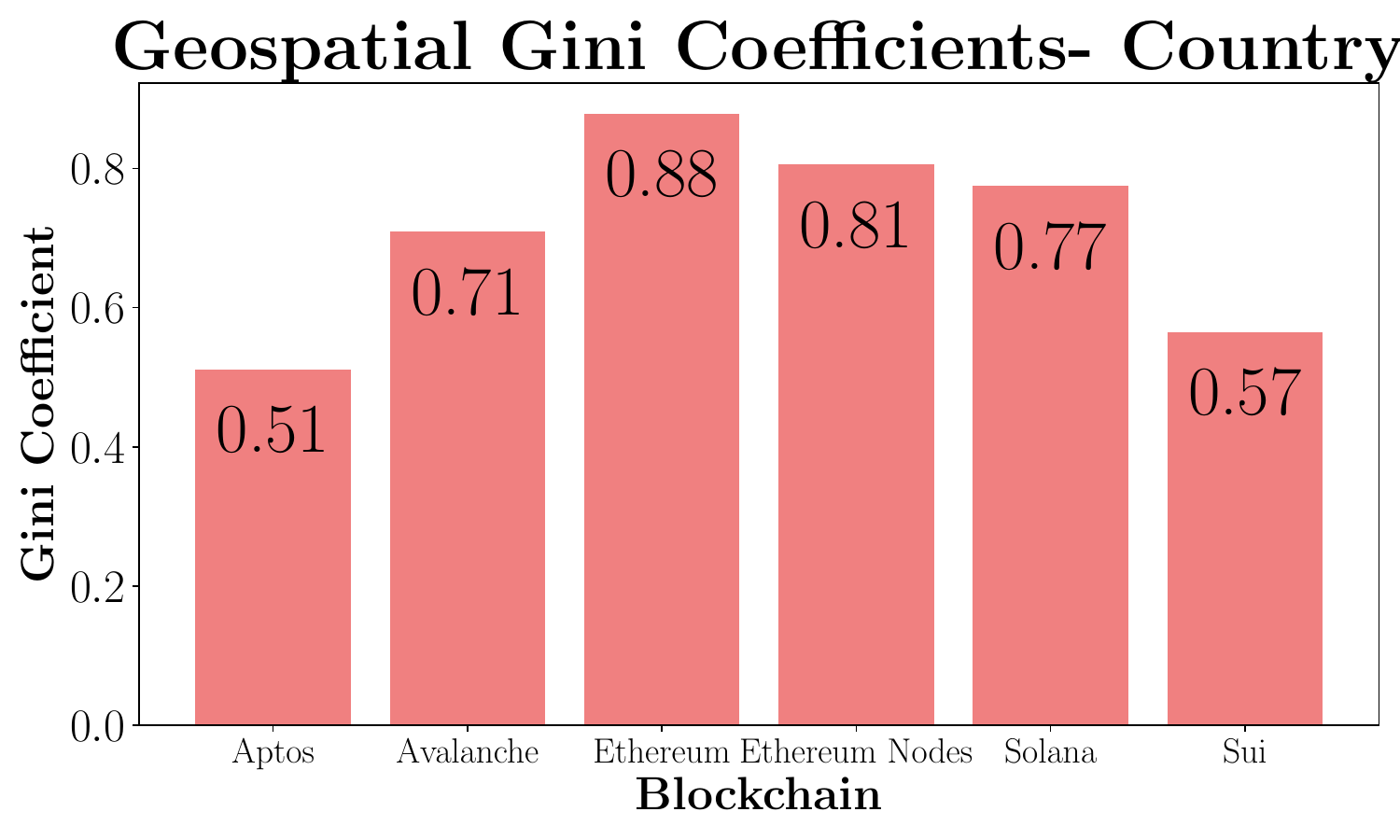} 
  \caption{Gini coefficients of various blockchains.}
  \label{fig:giniHistogram}
\end{figure}

\begin{figure}[htbp]
  \centering
  \includegraphics[width=\linewidth]{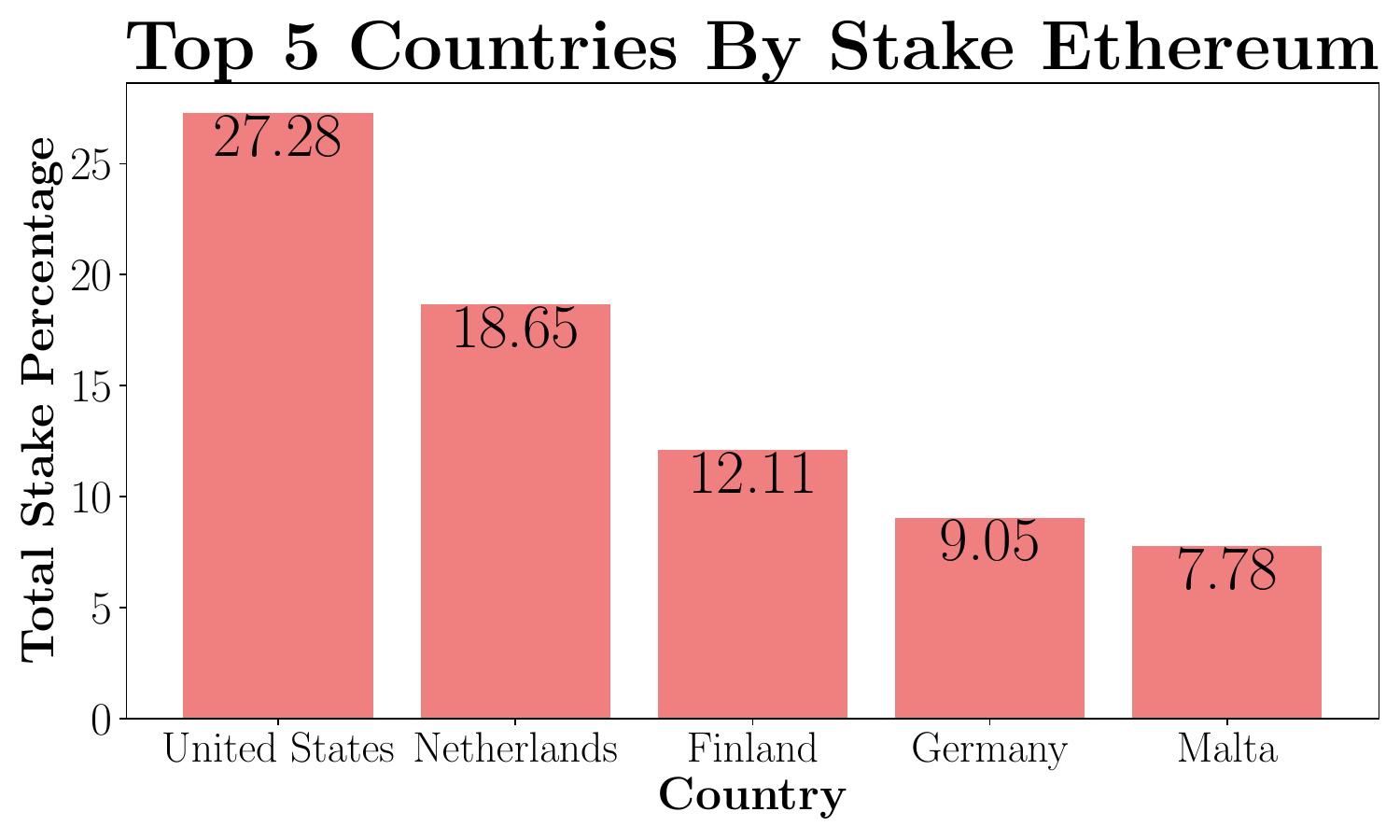}
  \caption{Stake distribution of Ethereum.}
  \label{fig:ethereumStakeHistogram}
\end{figure}

To determine the country of each validator, we utilized their geographical coordinates~\cite{opencage_geocoding_api} and subsequently aggregated the stake weights \(s_i\) by country. The aggregated stake \(S_c\) for a country \(c\) is expressed as:

\begin{equation}
S_c = \sum_{v_i \in C} s_i
\end{equation}

where \(C\) represents the set of validators located within country \(c\). Our analysis indicates that stake concentrations are significantly high in countries such as the United States, Germany, Finland, and the Netherlands across most blockchains. Figure~\ref{fig:ethereumStakeHistogram} presents the top five countries by aggregated stake for Ethereum, while tables in Appendix~\ref{appendix:top8} detail the top eight countries by aggregated stake for all blockchains. Notably, the top three countries account for over 33\% of the total stake across all blockchains, indicating a lack of geospatial decentralization and potential regulatory capture~\cite{censorshipEthereum}.

To quantify geospatial decentralization across blockchains, we utilize the Gini coefficient, a well-established metric for quantifying inequality~\cite{ceriani2012origins,gini1921measurement}. Mathematically, the Gini coefficient \( G \) is defined as:

\begin{equation}
G = \frac{\sum_{i=1}^n \sum_{j=1}^n | S_{c_i} - S_{c_j} |}{2n^2 \bar{S}}
\end{equation}

where \(S_{c_i}\) represents the aggregated stake of country \(c_i\), \(n\) is the number of countries, and \(\bar{S}\) is the mean aggregated stake across all countries. The Gini coefficient ranges from 0 (perfect equality) to 1 (maximal inequality), offering a clear metric to evaluate stake distribution and, consequently, the degree of geospatial decentralization.

In our analysis, all evaluated blockchains exhibit Gini coefficients exceeding 0.5, as illustrated in Figure~\ref{fig:giniHistogram}. This outcome signifies substantial centralization of stake. Specifically, Ethereum exhibits a Gini coefficient of 0.88, indicating a pronounced concentration of voting power within a few countries, undermining the principles of decentralized consensus.

While the Gini coefficient provides insights into decentralization at the country level, it does not capture variations within individual countries, given significant differences in their geographic sizes. To address this, we introduce a proximity-based Gini coefficient that aggregates stake within a specified radius around each validator. This metric allows us to quantify inequalities in stake distribution at a more granular level and highlights the lack of geospatial decentralization at regional scales. 

\clearpage
\subsection{Top 8 Countries by Stake Weight}
\label{appendix:top8}
\begin{table}[htbp]
    \centering
    \begin{tabular}{cc}
        \begin{minipage}{0.50\linewidth}
            \centering
            \caption{{Sui Validators Top 8 Countries}}
            \begin{tabular}{|l|c|}
            \hline
            Country & Stake Percentage \\
            \hline
            United States & 18.30 \\
            Germany & 13.76 \\
            United Kingdom & 10.62 \\
            Lithuania & 6.87 \\
            Netherlands & 6.77 \\
            France & 6.37 \\
            Japan & 6.07 \\
            Singapore & 4.80 \\
            \hline
            \end{tabular}
        \end{minipage} &
        \begin{minipage}{0.45\linewidth}
            \centering
            \caption{Ethereum Nodes Top 8 Countries}
            \begin{tabular}{|l|c|}
            \hline
            Country & Stake Percentage \\
            \hline
            United States & 30.71 \\
            Germany & 15.46 \\
            Finland & 4.79 \\
            United Kingdom & 4.28 \\
            France & 4.11 \\
            Netherlands & 3.61 \\
            Canada & 3.37 \\
            China & 3.15 \\
            \hline
            \end{tabular}
        \end{minipage} \\
        \begin{minipage}{0.45\linewidth}
           
            \caption{{Solana Validators Top 8 Countries}}
            \begin{tabular}{|l|c|}
            \hline
            Country & Stake Percentage \\
            \hline
            United States & 24.93 \\
            Germany & 15.23 \\
            Netherlands & 14.07 \\
            Japan & 9.12 \\
            United Kingdom & 8.10 \\
            France & 6.87 \\
            Lithuania & 5.54 \\
            Ireland & 2.87 \\
            \hline
            \end{tabular}
        \end{minipage} &
        \begin{minipage}{0.45\linewidth}
           
            \caption{{Aptos Validators Top 8 Countries}}
            \begin{tabular}{|l|c|}
            \hline
            Country & Stake Percentage \\
            \hline
            Germany & 12.07 \\
            United States & 11.47 \\
            Singapore & 10.30 \\
            Ireland & 9.16 \\
            Netherlands & 9.01 \\
            France & 8.87 \\
            South Korea & 8.05 \\
            United Kingdom & 7.09 \\
            \hline
            \end{tabular}
        \end{minipage} \\
        \begin{minipage}{0.45\linewidth}
            \centering
            \caption{{Ethereum Validators Top 8 Countries}}
            \begin{tabular}{|l|c|}
            \hline
            Country & Stake Percentage \\
            \hline
            United States & 27.28 \\
            Netherlands & 18.65 \\
            Finland & 12.11 \\
            Germany & 9.05 \\
            Malta & 7.78 \\
            France & 2.86 \\
            Canada & 2.53 \\
            Singapore & 2.37 \\
            \hline
            \end{tabular}
        \end{minipage} &
        \begin{minipage}{0.4\linewidth}
            \centering
            \caption{{Avalanche Validators Top 8 Countries}}
            \begin{tabular}{|l|c|}
            \hline
            Country & Stake Percentage \\
            \hline
            United States & 29.72 \\
            Germany & 14.14 \\
            Ireland & 8.32 \\
            Japan & 5.84 \\
            Singapore & 4.46 \\
            Canada & 4.45 \\
            Australia & 3.85 \\
            France & 3.61 \\
            \hline
            \end{tabular}
        \end{minipage}
    \end{tabular}
\end{table}
\clearpage
\section{Proximity-Based Gini Coefficient}
\label{appendix:gini-proximity-appendix}

To evaluate geospatial decentralization, we introduce the \textit{Proximity-Based Gini Coefficient}, which quantifies inequality in the aggregated stake of validators based on proximity rather than broader country-level groupings. This metric provides insights into localized stake distributions, revealing geospatial clustering.

We define a \textit{neighborhood} \(\mathcal{N}_i\) for each validator \(v_i\) as the set of validators within a distance threshold \(\Delta_{\text{threshold}}\):

\begin{equation}
    \mathcal{N}_i = \{v_j \in \mathcal{V} \mid \Delta_{ij} \leq \Delta_{\text{threshold}}, j \neq i\}
\end{equation}

The aggregated stake \(S_{\text{agg}, i}\) for each validator \(v_i\) is calculated as:

\begin{equation}
    S_{\text{agg}, i} = s_i + \sum_{v_j \in \mathcal{N}_i} s_j
\end{equation}

where \(s_i\) is the stake weight of validator \(v_i\). If a validator has no neighbors (\(\mathcal{N}_i = \emptyset\)), \(S_{\text{agg}, i} = s_i\).

The proximity-based Gini coefficient \(G_{\Delta}\) is computed over the set of aggregated stakes \(\{S_{\text{agg}, i}\}_{i=1}^n\):

\begin{equation}
    G_{\Delta} = \frac{\sum_{i=1}^{n} \sum_{j=1}^{n} |S_{\text{agg}, i} - S_{\text{agg}, j}|}{2n^2 \bar{S}_{\text{agg}}}
\end{equation}

where \(\bar{S}_{\text{agg}}\) is the mean of the aggregated stakes. This ensures \(G_{\Delta}\) ranges from 0 (complete equality) to 1 (maximum inequality). The proximity-based Gini highlights local inequalities that may be obscured by coarser metrics such as country-level Gini coefficients.

Table~\ref{table:proximity_gini} presents the proximity-based Gini coefficients for different distance thresholds, alongside the Gini coefficients based on country-level stake aggregation. Across all blockchains, the Gini values are consistently high, particularly at lower distance thresholds, with notable examples such as Ethereum exhibiting values of 0.88 at both country and 100 km scales. This indicates significant concentration of stake within limited geospatial regions, pointing to a lack of effective geospatial decentralization. Even for other blockchains like Solana and Avalanche, proximity-based Gini coefficients remain above 0.7 at smaller distances, suggesting that most influential validators are clustered geographically rather than being well-distributed. As distance thresholds increase, we observe a gradual decrease in Gini values, indicating minor improvements in geospatial diversity, but the persistence of relatively high Gini coefficients (\(>0.5\)) emphasizes that influence remains unevenly distributed, failing to achieve meaningful geographic spread. These findings highlight the critical need for mechanisms, such as GPoS, to enforce a more uniform distribution of validator stake and address regional clustering.

\begin{table}
\centering
\caption{Proximity-Based Gini Coefficients for Various Distance Thresholds}
\label{table:proximity_gini}
\renewcommand{\arraystretch}{1.5}
\setlength{\tabcolsep}{8pt}

\rotatebox{90}{
\resizebox{\textwidth}{0.2\textwidth}{%
\begin{tabular}{|l|l|l|l|l|l|l|l|}
\hline
  
\textbf{Blockchain} & \textbf{\begin{tabular}[c]{@{}l@{}}Gini by \\ country\end{tabular}} & \textbf{\begin{tabular}[c]{@{}l@{}}Gini by \\ proximity\\ 100km\end{tabular}} & \textbf{\begin{tabular}[c]{@{}l@{}}Gini by \\ proximity\\ 200km\end{tabular}} & \textbf{\begin{tabular}[c]{@{}l@{}}Gini by \\ proximity\\ 400km\end{tabular}} & \textbf{\begin{tabular}[c]{@{}l@{}}Gini by \\ proximity\\ 600km\end{tabular}} & \textbf{\begin{tabular}[c]{@{}l@{}}Gini by \\ proximity\\ 800km\end{tabular}} & \textbf{\begin{tabular}[c]{@{}l@{}}Gini by\\ proximity\\ 1000km\end{tabular}} \\ \hline
Aptos               & 0.51                                                                & 0.57                                                                          & 0.57                                                                          & 0.61                                                                          & 0.62                                                                          & 0.61                                                                          & 0.58                                                                          \\ \hline
Avalanche           & 0.71                                                                & 0.71                                                                          & 0.65                                                                          & 0.55                                                                          & 0.51                                                                          & 0.47                                                                          & 0.43                                                                          \\ \hline
Ethereum            & 0.88                                                                & 0.88                                                                          & 0.82                                                                          & 0.72                                                                          & 0.64                                                                          & 0.58                                                                          & 0.52                                                                          \\ \hline
Ethereum nodes      & 0.81                                                                & 0.72                                                                          & 0.66                                                                          & 0.62                                                                          & 0.58                                                                          & 0.54                                                                          & 0.49                                                                          \\ \hline
Solana              & 0.77                                                                & 0.76                                                                          & 0.72                                                                          & 0.72                                                                          & 0.66                                                                          & 0.61                                                                          & 0.55                                                                          \\ \hline
Sui                 & 0.57                                                                & 0.53                                                                          & 0.58                                                                          & 0.58                                                                          & 0.58                                                                          & 0.55                                                                          & 0.50                                                                          \\ \hline
\end{tabular}
}}
\end{table}
\clearpage
\section{GPoS Evaluation: Gini by Country}
\label{appendix:gpos-evaluation}

\begin{figure}[ht]
    \centering
    \includegraphics[width=\textwidth, trim=10 0 2 10, clip]{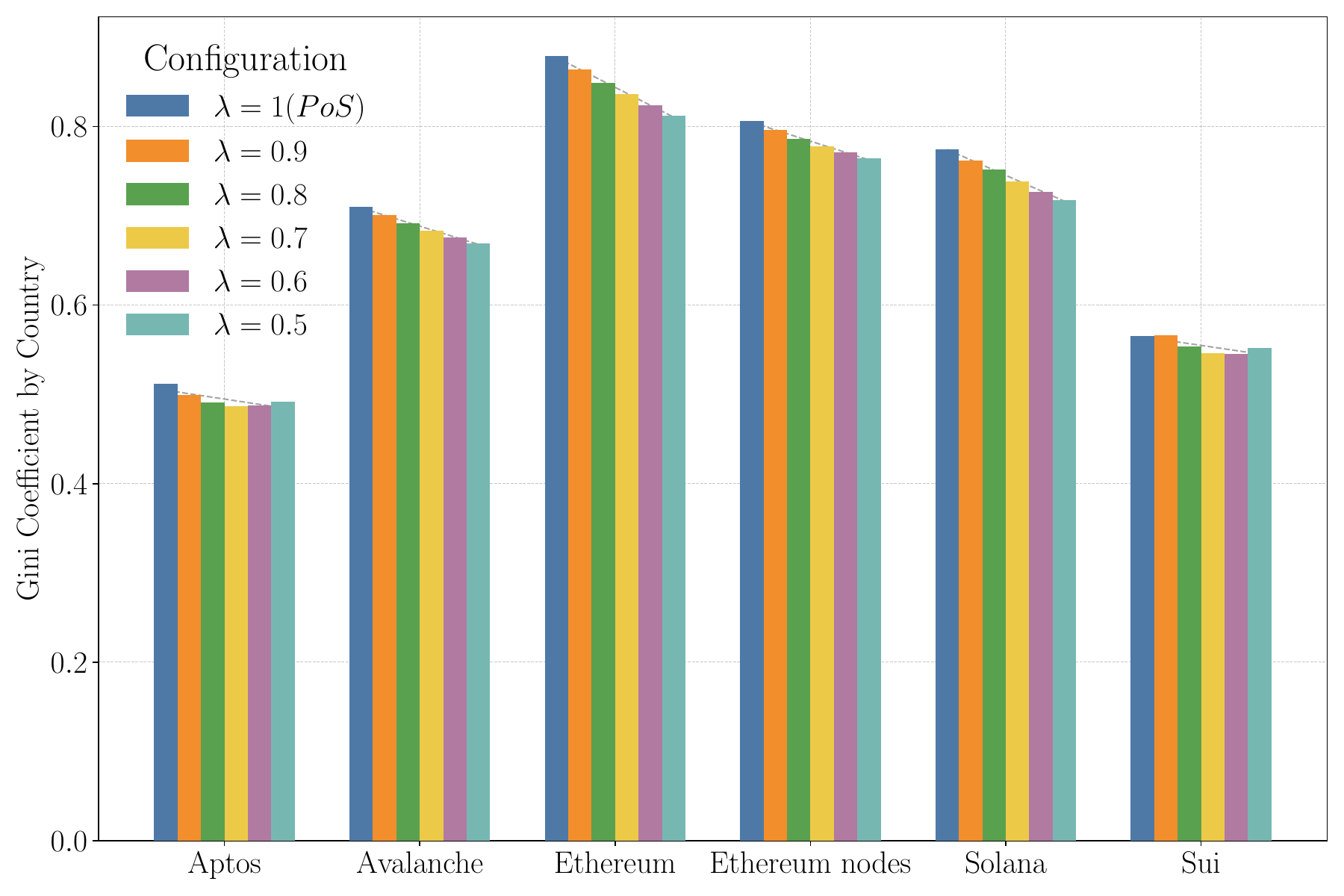} 
    \caption{Gini coefficients for voting power aggregated by country, with varying \( \lambda \) values.}
    \label{fig:gini_country_gpos}
\end{figure}

\section{Exponential GPoS}
\label{sub-appendix:exponential}

In addition to the linear combination model of GPoS introduced in Section~\ref{subsec:votingpowe}, we explored an alternative exponential formulation. In this model, the voting power of a validator \( v_i \) is defined as:

\begin{equation}
    \rho_i = s_i^{(\alpha)} \cdot GDI_i^{(1 - \alpha)}
\end{equation}

where \( \alpha \in [0, 1] \) is a tunable parameter that controls the balance between  \( s_i \) and \( GDI_i \). Both \( s_i \) and \( GDI_i \) should be normalized within the interval \([0, 1]\) to ensure neither variables disproportionately influence the voting power.

In this formulation, \(\alpha\) determines the relative weight of stake versus GDI. When \(\alpha = 1\), the model reduces to traditional PoS, with voting power based solely on stake. As \(\alpha\) decreases, GDI contributes more significantly to voting power, enhancing geospatial diversity.

Figure~\ref{fig:gini_country_gpos} presents the Gini coefficients for eigenvector centrality scores computed using the exponential model. Results indicate a consistent decline in Gini values as \(\alpha\) decreases, highlighting the benefits of incorporating geospatial diversity into consensus mechanisms. On average, the Gini coefficients decreased by 30\% as \(\alpha\) moved from 1 to 0.5, demonstrating the effectiveness of this approach in mitigating influence centralization.

\begin{figure}[ht]
    \centering
    \includegraphics[width=\textwidth, trim=10 0 2 10, clip]{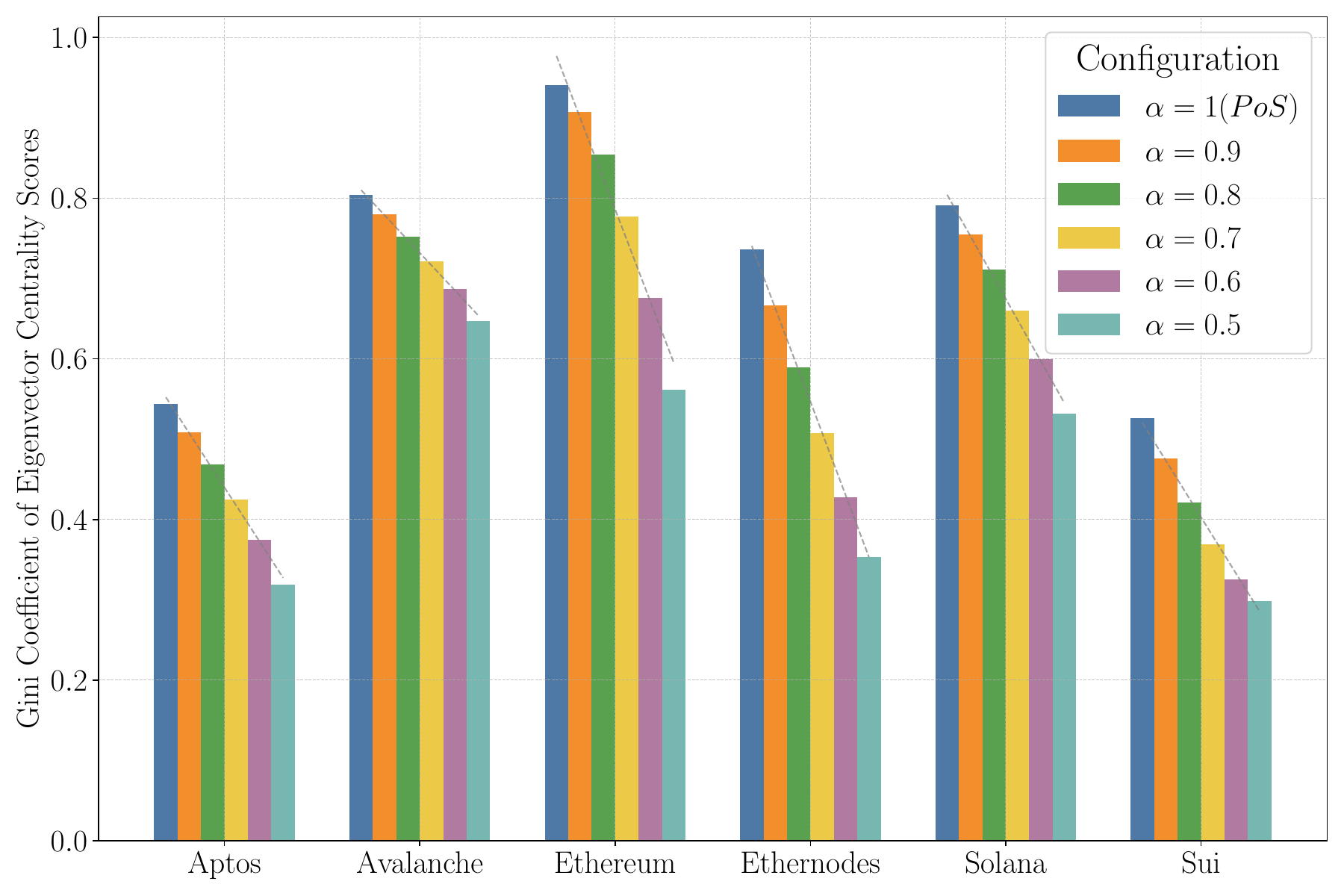} 
    \caption{Gini coefficients for eigenvector centrality measure, with varying \( \alpha \) values in expontial setting.}
    \label{fig:gini_centrality_exponential}
\end{figure}

\end{document}